\documentclass[useAMS,usenatbib,letterpaper]{mn2e}

\usepackage{times}
\usepackage{psfrag}
\usepackage[dvips]{graphicx}
\usepackage{amssymb}
\usepackage{bm}
\usepackage[french,english]{babel}
\usepackage{color}

\title[String signal reconstruction]
{Compressed sensing reconstruction of a string signal from interferometric observations of the cosmic microwave background}

\author[Wiaux et al.]
{Y. Wiaux$^{1,2}$\thanks{E-mail: yves.wiaux@epfl.ch}, G. Puy$^{1}$, P. Vandergheynst$^{1}$\\
$^{1}$Institute of Electrical Engineering, Ecole Polytechnique F\'ed\'erale de Lausanne (EPFL), CH-1015 Lausanne, Switzerland\\
$^{2}$Centre for Particle Physics and Phenomenology, Universit\'e catholique de Louvain (UCL), B-1348 Louvain-la-Neuve, Belgium\\}

\begin{document}

\date{\today}

\pagerange{\pageref{firstpage}--\pageref{lastpage}} \pubyear{2009}

\maketitle

\label{firstpage}

\begin{abstract}

We propose an algorithm for the reconstruction of the signal induced by cosmic strings in the cosmic microwave background (${\rm CMB}$), from radio-interferometric data at arcminute resolution. Radio interferometry provides incomplete and noisy Fourier measurements of the string signal, which exhibits sparse or compressible magnitude of the gradient due to the Kaiser-Stebbins (${\rm KS}$) effect. In this context the versatile framework of compressed sensing naturally applies for solving the corresponding inverse problem. Our algorithm notably takes advantage of a model of the prior statistical distribution of the signal fitted on the basis of realistic simulations. Enhanced performance relative to the standard ${ \rm CLEAN}$ algorithm is demonstrated by simulated observations under noise conditions including primary and secondary ${\rm CMB}$ anisotropies.

\end{abstract}

\begin{keywords}
techniques: image processing -- techniques: interferometric -- cosmic microwave background.
\end{keywords}
\section{Introduction}\label{sec:Introduction}
Recent observations of the cosmic microwave background (${\rm CMB}$) and of the Large Scale Structure of the Universe (${\rm LSS}$) have led to the definition of a concordance cosmological model. The analysis of the ${\rm CMB}$ temperature data over the whole celestial sphere from the Wilkinson Microwave Anisotropy Probe (${\rm WMAP}$) satellite experiment has been playing a dominant role in designing this precise picture of the Universe \citep{bennett03,spergel03,hinshaw07,spergel07,hinshaw09,komatsu09}. According to this model, the cosmic structures and the ${\rm CMB}$ originate from Gaussian adiabatic perturbations seeded in the early phase of inflation of the Universe. However, cosmological scenarios motivated by theories for the unification of the fundamental interactions predict the existence of topological defects, resulting from phase transitions at the end of inflation \citep{vilenkin94,hindmarsh95a,hindmarsh95b,turok90}. These defects would have participated to the formation of the cosmic structures, also imprinting the ${\rm CMB}$. Cosmic strings are a line-like version of such defects which are also predicted in the framework of fundamental string theory \citep{davis05}. As a consequence, the issue of their existence is a central question in cosmology today.

Cosmic strings are parametrized by a string tension $\mu$, i.e. a mass per unit length, which sets the overall strength of a string network. Their main signature in the ${\rm CMB}$ is characterized by temperature steps along the string positions. This localized effect, known as the Kaiser-Stebbins (${\rm KS}$) effect \citep{kaiser84,gott85}, hence implies a non-Gaussian imprint of the string network in the ${\rm CMB}$. The most numerous strings appear at an angular size around $1$ ${\rm deg}$ on the celestial sphere. ${\rm CMB}$ experiments with an angular resolution much below $1$ ${\rm deg}$ are thus required in order to resolve the width of cosmic strings.

Experimental constraints have been obtained on a possible string contribution in terms of upper bounds on the string tension $\mu$ \citep{perivolaropoulos93,bevis04,wyman05,wyman06,bevis07,fraisse07}. In this context, even though observations largely fit with an origin of the cosmic structures in terms of adiabatic perturbations, room is still available for the existence of cosmic strings.

The observed ${\rm CMB}$ signal can be modeled as a linear superposition of a statistically isotropic but non-Gaussian string signal $x$ proportional to an unknown string tension, with statistically isotropic Gaussian noise $g$ comprising the standard components of the ${\rm CMB}$ induced by adiabatic perturbations, as well as instrumental noise. In a first recent work \citep{hammond09} we developed an effective method for mapping the string network potentially imprinted in ${\rm CMB}$ temperature data, in the perspective of forthcoming arcminute resolution experiments performing a mapping of planar patches of the sky in real space, with standard radiometers. We took a Bayesian approach to this denoising problem, based on statistical models for both the string signal and the noise. The denoising is done in the wavelet domain, using a steerable wavelet transform well adapted for representing the strongly oriented features present in the string signal. We showed that the string signal coefficients are well modeled by generalized Gaussian distributions (${\rm GGD}$'s), which are fitted at each wavelet scale using a training simulation borrowed from the set of realistic string simulations recently produced by \citet{fraisse08}.

In the present work we aim at extending the approach described above to the reconstruction of a string signal from radio-interferometric observations of the ${\rm CMB}$ probing small patches of the celestial sphere at the same arcminute resolution. Under the most standard assumptions, radio interferometers provide a noisy and incomplete Fourier coverage of the planar signal. In these terms, the signal reconstruction amounts to solving an ill-posed inverse problem encompassing both denoising and deconvolution. Relying on the ${\rm KS}$ effect in the ${\rm CMB}$, the string network itself can be mapped by the magnitude of the gradient $\nabla x$ of the string signal, which is very sparse or compressible in nature, with significant amplitudes only along the strings themselves.

The theory of compressed sensing offers a new framework for solving ill-posed inverse problem when the signals are sparse or compressible \citep{candes06a,candes06b,candes06,donoho06,baraniuk07a,donoho09}. A band-limited signal may be expressed as the $N^2$-dimensional vector of its values sampled at the Nyquist-Shannon rate. By definition, a signal is sparse in some basis if its expansion contains only a small number $K\ll N^2$ of non-zero coefficients. More generally it is compressible if its expansion only contains a small number of significant coefficients, i.e. if a large number of its coefficients bear a negligible value. The theory of compressed sensing demonstrates that, for sparse or compressible signals, a small number $M\ll N^2$ of random measurements, in a sensing basis incoherent with the sparsity basis, will suffice for an accurate and stable reconstruction of the signals. The mutual coherence between two bases may be defined as the maximum complex modulus of the scalar product between unit-norm vectors of the two bases. A random selection of Fourier measurements of a signal sparse in real space or in its gradient are particular examples of good sensing procedures. In this framework, the reconstruction of the signals may be approached in different ways, but all of them encompass a regularization of the ill-posed inverse problem by the introduction of a sparsity or compressibility constraint. In a second recent work \citep{wiaux09a}, we presented results showing that compressed sensing offers powerful image reconstruction techniques for radio-interferometric data, which are stable relative to noise, as well as stable relative to non-exact sparsity, i.e. compressibility of the signals. These techniques are based on global minimization problems, which are solved by convex optimization algorithms \citep{combettes07,vandenBerg08}. We particularly illustrated the versatility of the scheme relative to the inclusion of specific prior information on the signal in the minimization problems.

We here wish to apply these results for the reconstruction of a string signal in the ${\rm CMB}$. As for the denoising algorithm proposed in \citet{hammond09}, the reconstruction algorithm that we present may be considered as a modular component of a larger data analysis. It may notably find use as a pre-processing step for other methods for cosmic string detection based on explicit edge detection \citep{jeong05,lo05,amsel07,danos08}. Our reconstruction algorithm is tested under different conditions with astrophysical noise components including various contributions to the standard components of the ${\rm CMB}$, i.e. primary and secondary anisotropies. Our analyses rely on the set of realistic string simulations recently produced by \citet{fraisse08}.

The remainder of this paper is organized as follows. In Section \ref{sec:String signal and noise}, we discuss the string signal and noise in the ${\rm CMB}$. In Section \ref{sec:Interferometric inverse problem}, we describe the inverse problem posed in the context of radio interferometry. In Section \ref{sec:Compressed sensing reconstruction}, we define the minimization problems to be solved for signal reconstruction in the context of compressed sensing, notably relying on a prior statistical model of the string signal fitted from simulations. In Section \ref{sec:Analyses and results}, we describe our numerical analyses and results. We finally conclude in Section \ref{sec:Conclusion}.
\section{String signal and noise}\label{sec:String signal and noise}
In this section, we describe the string signal in the ${\rm CMB}$, as well as the astrophysical noise made up of the standard primary and secondary anisotropies. We also describe the numerical simulations used for our subsequent analyses.
\subsection{String signal}
In an inflationary cosmological model, the phase transitions responsible for the formation of a cosmic string network occur after the end of inflation, so as to produce observable defects.  From the epoch of last scattering until today, the cosmic string network
continuously imprints the ${\rm CMB}$. The so-called scaling solution for the string network implies that the most numerous strings are imprinted just after last scattering and have a typical angular size around $1$ ${\rm deg}$, of the order of the horizon size at that time \citep{vachaspati84, kibble85,albrecht85,bennet86,bennet89,albrecht89,bennet90,allen90}. Longer strings are also imprinted in the later stages of the Universe evolution, but in smaller number, according to the number of corresponding horizon volumes required to fill the sky.

The main signature of a cosmic string in the ${\rm CMB}$ is described by the ${\rm KS}$ effect according to which a temperature step is induced along the string position. The relative amplitude of this step is given by
\begin{equation}
\frac{\delta T}{T}=\left(8\pi\gamma\beta\right)\rho,\label{eq:ssn1}
\end{equation}
where $\beta=v/c$ and $\gamma=(1-\beta^{2})^{-1/2}$ with $v$ standing for the string velocity transverse to the line of sight and $c$ for the speed of light, and where $\rho$ is a dimensionless parameter uniquely associated with the string tension $\mu$ through
\begin{equation}
\rho=\frac{G\mu}{c^{2}},\label{eq:ssn2}
\end{equation}
with $G$ standing for the gravitational constant. In the following we call $\rho$ the string tension.

On small angular scales, realistic simulations can be produced by stacking ${\rm CMB}$ maps induced in different redshift ranges between last scattering and today. The simulations that we use in this work have been produced by this technique \citep{bouchet88,fraisse08}. The string signal is understood as a realization of a statistically isotropic but non-Gaussian process on the celestial sphere with an overall amplitude rescaled by the string tension $\rho$, and characterized by a nearly scale-free angular power spectrum: $C_{l}^{(x)}(\rho)=\rho^{2}C_{l}^{(x)}$, where the positive integer index $l$ stands for the angular frequency index on the sphere.  An analytical expression of this spectrum was provided for $l$ larger than a few hundreds by \citet{fraisse08}, on the basis of their simulations.

We consider radio-interferometric ${\rm CMB}$ observations with a small field of view corresponding to an angular opening $L\in[0,2\pi)$ on the celestial sphere. In this context, the small portion of the celestial sphere accessible is identified to a planar patch of size $L\times L$, and we may consider planar signals functions of a two-dimensional position vector $\bm{p}$. The corresponding spatial frequencies may be denoted as two-dimensional vectors $\bm{k}$ with a radial component given by the norm $k$ of the vector.  In this Euclidean limit, the radial component identifies with the angular frequency on the celestial sphere, below some band limit $B$ set by the resolution of the interferometer under consideration: $l=k<B$.  For $k$ larger than a few hundreds, the nearly scale-free planar power spectrum of the string signal $x(\bm{p})$ reads as:
\begin{equation}
 P^{(x)}\left(k,\rho\right)=\rho^{2}P^{(x)}\left(k\right),\label{eq:ssn3}
\end{equation}
with $P^{(x)}(k)=C_{l}^{(x)}$ for $l=k$.

The observed ${\rm CMB}$ signal can be understood as a linear superposition of the string signal and statistically isotropic noise of astrophysical and instrumental origin $g(\bm{p})$ with some angular power spectrum $C_{l}^{(g)}$. In the Euclidean limit, the corresponding planar power spectrum for the noise $g(\bm{p})$ may be written as $P^{(g)}(k)=C_{l}^{(g)}$ for $l=k$. Let us note that we consider zero mean signals, identifying perturbations around statistical means.
\subsection{Astrophysical noise}
\begin{figure}
\begin{center}
\psfrag{1e+04}{\hspace{0.2cm}{\scriptsize $10^4$}}
\psfrag{1e+03}{\hspace{0.2cm}{\scriptsize $10^3$}}
\psfrag{1e+02}{\hspace{0.2cm}{\scriptsize $10^2$}}
\psfrag{1e+00}{\hspace{0.2cm}{\scriptsize $1$}}
\psfrag{1e-02}{\hspace{0.1cm}{\scriptsize $10^{-2}$}}
\psfrag{tagx}{\hspace{0cm}{\scriptsize $l$}}
\psfrag{tagy}{\hspace{-0.6cm}{\scriptsize $l(l+1)C_l/2\pi$ ($\mu{\rm K}^2$)}}

\includegraphics[width=8cm,keepaspectratio]{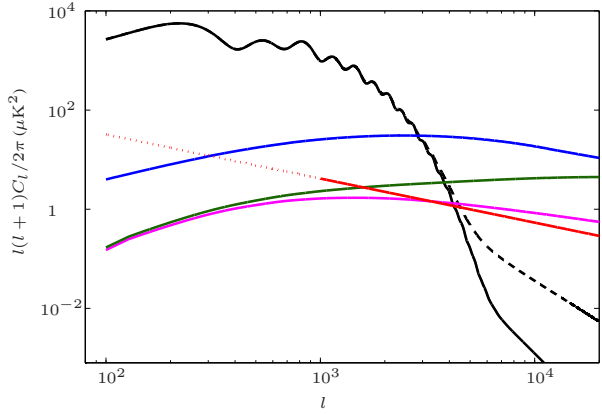}
\end{center}
\caption{\label{fig:spectra} Angular power spectra of the string signal and noise (borrowed from \citet{fraisse08}) as functions of the angular frequency in the range $l\in[10^2,2\times10^{4}]$ in $\log_{10}-\log_{10}$ axes scaling. The spectrum of the string signal is represented for a string tension $\rho=2\times10^{-7}$ in terms of its analytical expression valid at high angular frequencies (red straight line). The noise spectra (ordered by decreasing amplitude at low angular frequencies) are: the primary ${\rm CMB}$ anisotropies (black solid line) and the gravitational lensing correction (black dashed line), the thermal ${\rm SZ}$ effect in the Rayleigh-Jeans limit (blue solid line), the non-linear kinetic ${\rm SZ}$ effect (green solid line), and the Ostriker-Vishniac effect (magenta solid line).}
\end{figure}

Current ${\rm CMB}$ experiments achieve an angular resolution on the celestial sphere of the order of $10$ arcminutes, corresponding to a limit angular frequency not far above $B\simeq10^{3}$. At such resolutions, the standard components of the ${\rm CMB}$ primarily contain the statistically isotropic and Gaussian primary anisotropies induced by adiabatic perturbations at last scattering. In this context, any possible string signal is confined to amplitudes largely dominated by these primary anisotropies. The constraints mainly come from a best fit analysis of the angular power spectrum of the overall ${\rm CMB}$ signal in the ${\rm WMAP}$ temperature data \citep{perivolaropoulos93,bevis04,wyman05,wyman06,bevis07,fraisse07}. The tightest of these constraints \citep{fraisse07} gives the following upper bound at $68$ per cent confidence level:
\begin{equation}
\rho\leq\rho^{{\rm (exp)}}=2.1 \times 10^{-7} \label{eq:ssn4}.
\end{equation}

Forthcoming experiments will provide access to higher angular resolution. Among other instruments, radio interferometers such as the Arcminute Microkelvin Imager (AMI) will probe the ${\rm CMB}$ at angular resolutions below $1$ ${\rm arcmin}$ \citep{jones02,barker06,zwart08}. At these resolutions, the secondary ${\rm CMB}$ anisotropies induced by interaction of ${\rm CMB}$ photons with the evolving universe after last scattering will dominate the primary anisotropies, and must be accounted for.

The primary anisotropies exhibit exponential damping at high angular frequencies. This contrasts with the slow decay of the nearly scale-free angular power spectrum of the string signal, which thus dominates over the primary anisotropies at high enough angular frequencies. The secondary anisotropies include gravitational effects such as the Integrated Sachs-Wolfe (${\rm ISW}$) effect, the Rees-Sciama (${\rm RS}$) effect, and gravitational lensing, as well as re-scattering effects such as the thermal and kinetic Sunyaev-Zel'dovich (${\rm SZ}$) effects. The ${\rm SZ}$ effects dominate these secondary anisotropies \citep{sunyaev80,komatsu02,fraisse08}. The  ${\rm ISW}$  and  ${\rm RS}$  effects associated with the time evolution of the standard gravitational potentials can be neglected at these angular frequencies.  One may thus restrict the secondary anisotropies considered in the noise to the linear (Ostriker-Vishniac) and non-linear kinetic ${\rm SZ}$ effects, as well as the thermal ${\rm SZ}$ effect. The ${\rm SZ}$ effects are actually non-Gaussian, spatially dependent, and the kinetic and thermal effects are correlated.  As a simplifying assumption, we treat these two effects as two independent statistically isotropic Gaussian noise components. The effect of gravitational lensing is very small relative to the ${\rm SZ}$ effects, but we still take it into account as a correction to the angular power spectrum of the primary anisotropies.

At arcminute resolution, the thermal and kinetic ${\rm SZ}$ effects have standard deviations around $10$ and $5$ $\mu{\rm K}$ respectively. They also have a slow decay at high angular frequencies and will dominate the string signal for string tension values below the current experimental upper bound. Arcminute ${\rm CMB}$ experiments are in fact primarily dedicated to the detection of these secondary anisotropies. Unlike the other effects considered, which have the same black body spectrum as the primary anisotropies, the thermal ${\rm SZ}$ effect on the ${\rm CMB}$ temperature depends on the frequency of observation. Its amplitude decreases between the Rayleigh-Jeans limit (null frequency) and $217$ ${\rm GHz}$ where it is expected to vanish, before increasing again at higher frequencies. Figure \ref{fig:spectra} represents the angular power spectra  as functions of the angular frequency $l$, for a string signal with string tension $\rho=2\times10^{-7}$, for the primary ${\rm CMB}$ anisotropies and the correction due to gravitational lensing, for the Ostriker-Vishniac and the non-linear kinetic ${\rm SZ}$ effect, and for the thermal ${\rm SZ}$ effect in the Rayleigh-Jeans limit. These spectra are explicitly borrowed from \citet{fraisse08}.

In this context, the performance of the reconstruction algorithm to be defined will be studied in the following limits.  In a first noise condition, we will consider the secondary anisotropies as a statistically isotropic Gaussian noise, with power spectrum given by the Rayleigh-Jeans limit, that is added to the primary anisotropies. In a second noise condition, we will also assume an observation frequency around $217$ ${\rm GHz}$ taking advantage of the frequency dependence of the thermal ${\rm SZ}$ effect, and include in the noise secondary anisotropies in absence of this effect. This is equivalent to including only the kinetic ${\rm SZ}$ effect and gravitational lensing in the secondary anisotropies.  Instrumental noise is considered to be negligible and simply discarded. These two different noise conditions are respectively denoted as ${\rm SA+tSZ}$ (secondary anisotropies with thermal ${\rm SZ}$ effect) and ${\rm SA-tSZ}$ (secondary anisotropies without thermal ${\rm SZ}$ effect) in the following. Analyzing these limits can reveal to what extent the kinetic and thermal ${\rm SZ}$ effects hamper the  reconstruction of the string signal, as a function of the string tension. In a third noise condition, the reconstruction performance will also be examined in the limit where the noise only includes primary anisotropies, neglecting even instrumental noise, and assuming secondary anisotropies have been correctly separated. This case is denoted as ${\rm PA-IN}$ (primary anisotropies without instrumental noise) and will be studied in order to understand the behaviour of the reconstruction algorithm in ideal conditions.

For the sake of our analyses, foreground emissions such as Galactic dust or point sources \citep{kosowsky06} are disregarded.
\subsection{Numerical simulations}\label{sub:Numerical simulations}
\begin{figure}
\begin{center}
\includegraphics[width=4cm,keepaspectratio]{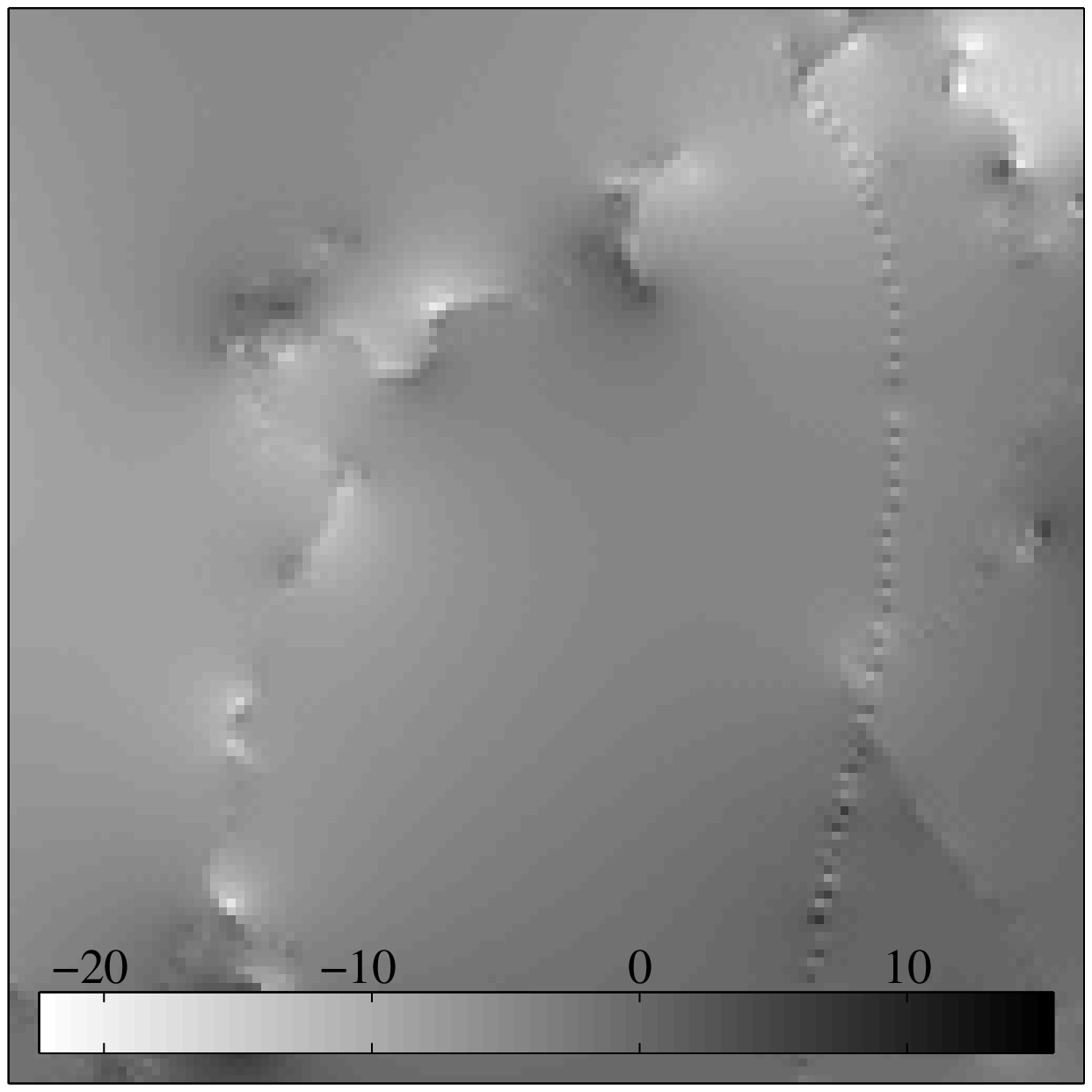}
\hspace{0.2cm}
\includegraphics[width=4cm,keepaspectratio]{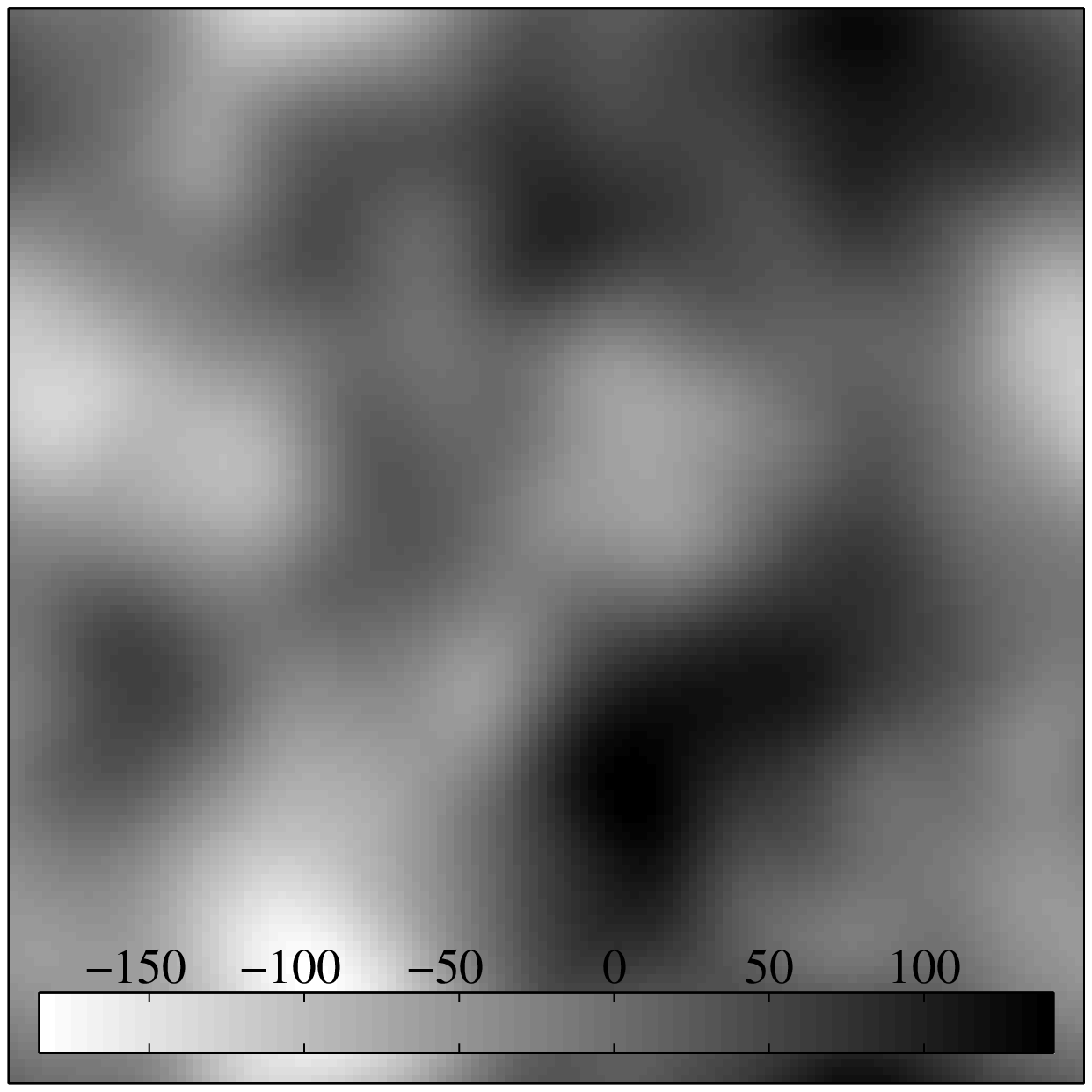}\\
\includegraphics[width=4cm,keepaspectratio]{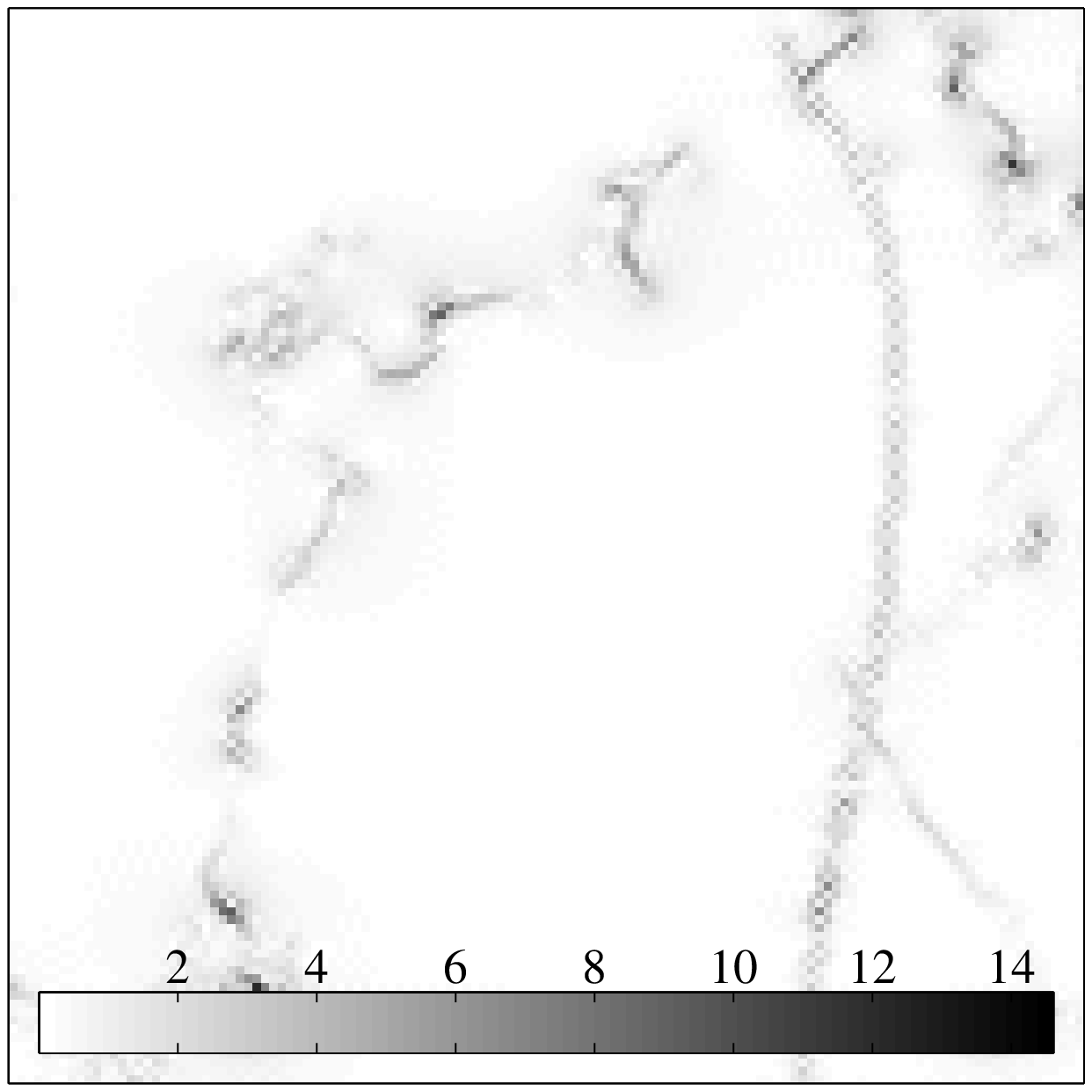}
\hspace{0.2cm}
\includegraphics[width=4cm,keepaspectratio]{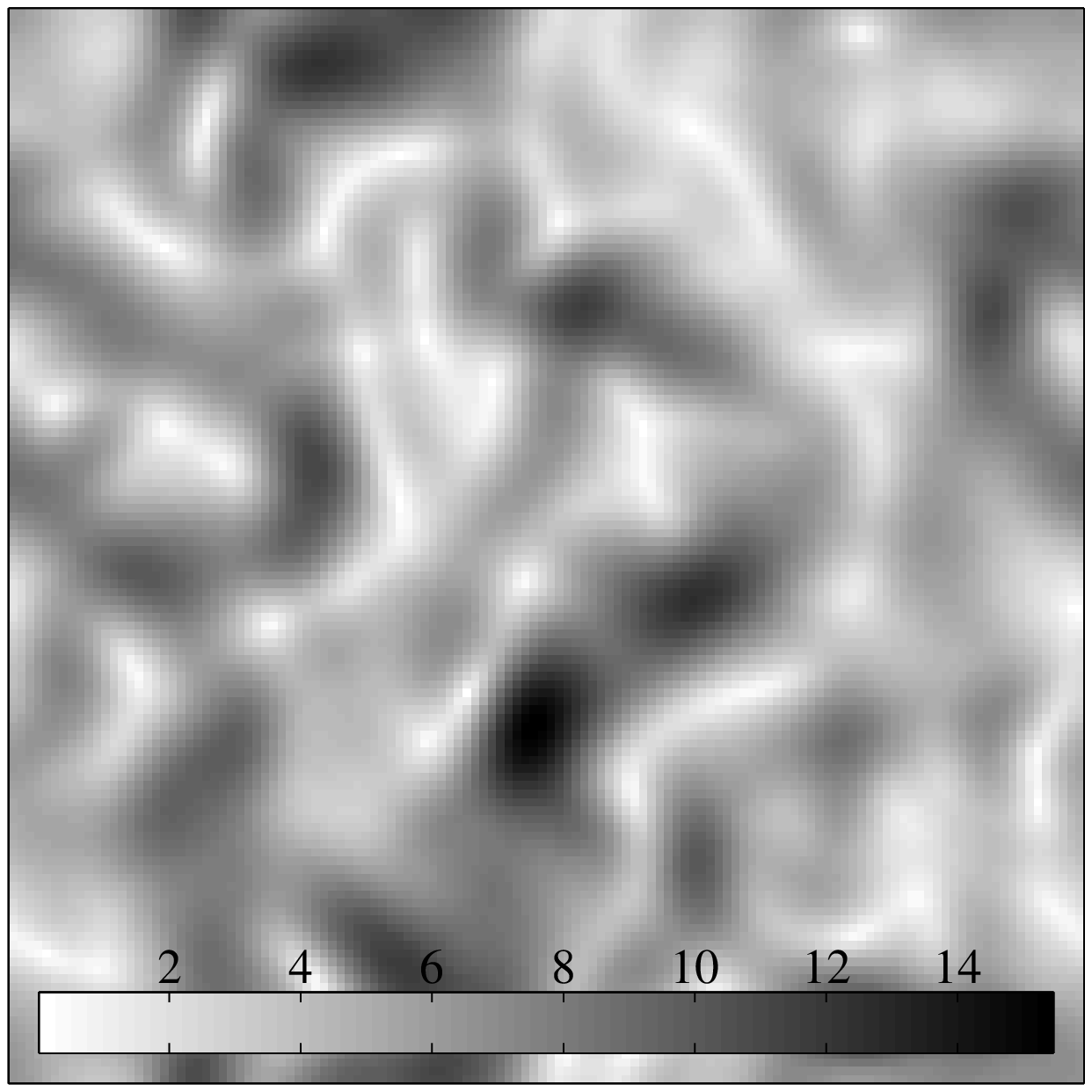}
\end{center}
\caption{\label{fig:maps-gradients} Simulated temperature maps (in ${\rm \mu K}$) of the string signal in the ${\rm CMB}$ and of its magnitude of gradient for a tension $\rho= 2\times 10^{-7}$ are represented in the top and bottom left panels respectively. Temperature maps of the primary ${\rm CMB}$ anisotropies, i.e. the noise in the ${\rm PA-IN}$ condition, and of its magnitude of gradient are represented in the top and bottom right panels respectively.}
\end{figure}

We use simulations of the string signal borrowed from the set of realistic simulations produced by \citep{fraisse08}. The specific simulations considered are defined on planar patches of size $L\times L$ for a field of view defined by an angular opening $L=0.9$ ${\rm deg}$. The finite size of the patch induces a discretization of the spatial frequencies which may be defined on a grid with $N\times N$ uniformly sampled points $\bm{k}_i$ with $1\leq i\leq N^{2}$, below some band limit $B$. The original maps are sampled at Nyquist-Shannon rate on grids with $N\times N$ uniformly sampled points $\bm{p}_{i}$ with $1\leq i\leq N^{2}$. We consider here samples with $N=128$, so that the corresponding pixels have an angular size around $0.42'$, with a band limit $B\simeq2.5\times10^{4}$ on each each component of the spatial frequencies.

For each noise component, a simulation may easily be produced by taking the Fourier transform of Gaussian white noise, renormalizing each Fourier coefficient by the square root of the corresponding power spectrum, and inverting the Fourier transform \citep{rocha05}. In each noise condition considered, an overall noise simulation is obtained by simple superposition of the required independent components simulated. The power spectrum of the noise $P^{(g)}(k)$ is the sum of the individual spectra.

Our approach for signal reconstruction notably relies on the introduction of specific statistical prior information on the signal in the regularization of the ill-posed inverse problem. A number of $64$ simulations of the string signal $x$ are used as training data for fitting a prior statistical model for the gradient of the signal. Only $1$ additional simulation is reserved for testing the algorithm. In all noise conditions considered and for each string tension, this test string signal simulation is combined with $30$ independent realizations of the noise in order to produce multiple test simulations. For illustration, Figure \ref{fig:maps-gradients} represents the test simulation of the string signal and one simulation of the noise, as well as corresponding maps of the magnitude of gradient.

Note that due to the finite size of the simulations, different string signal simulations exhibit slightly different estimated power spectra. For simplicity of our argument though both training and test simulations have been normalized so as to exhibit the same total power for the same string tension (see \citet{hammond09} for more details). 
\section{Interferometric inverse problem}\label{sec:Interferometric inverse problem}
In this section, we describe the ill-posed inverse problem posed for reconstruction of the string signal in the context of radio-interferometric observations of the ${\rm CMB}$. We also describe a modified inverse problem including a whitening operation on the measured visibilities and discuss basic ideas for signal reconstruction.
\subsection{Inverse problem}
\begin{figure}
\begin{center}
\includegraphics[width=4cm,keepaspectratio]{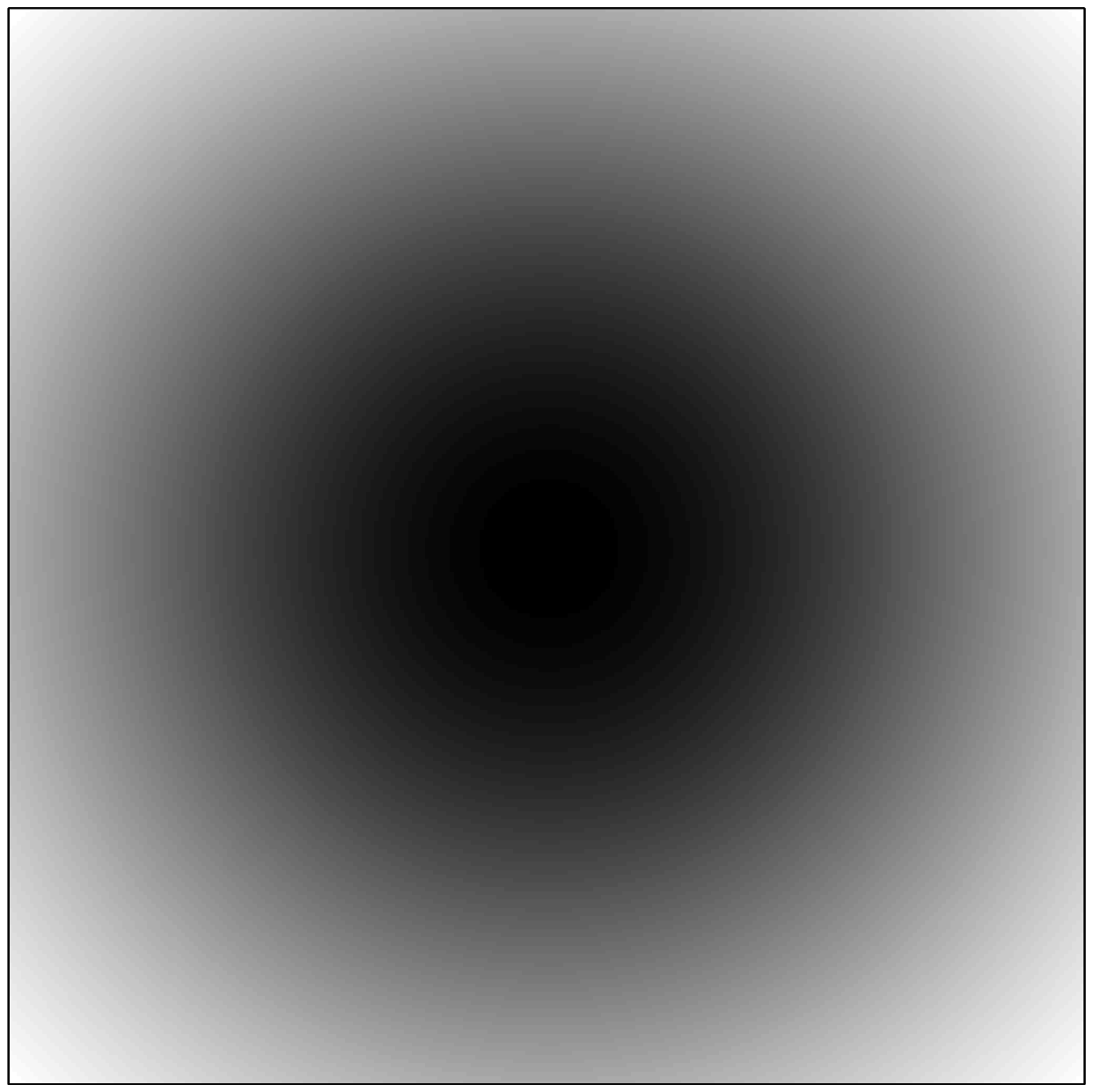}
\hspace{0.2cm}
\includegraphics[width=4cm,keepaspectratio]{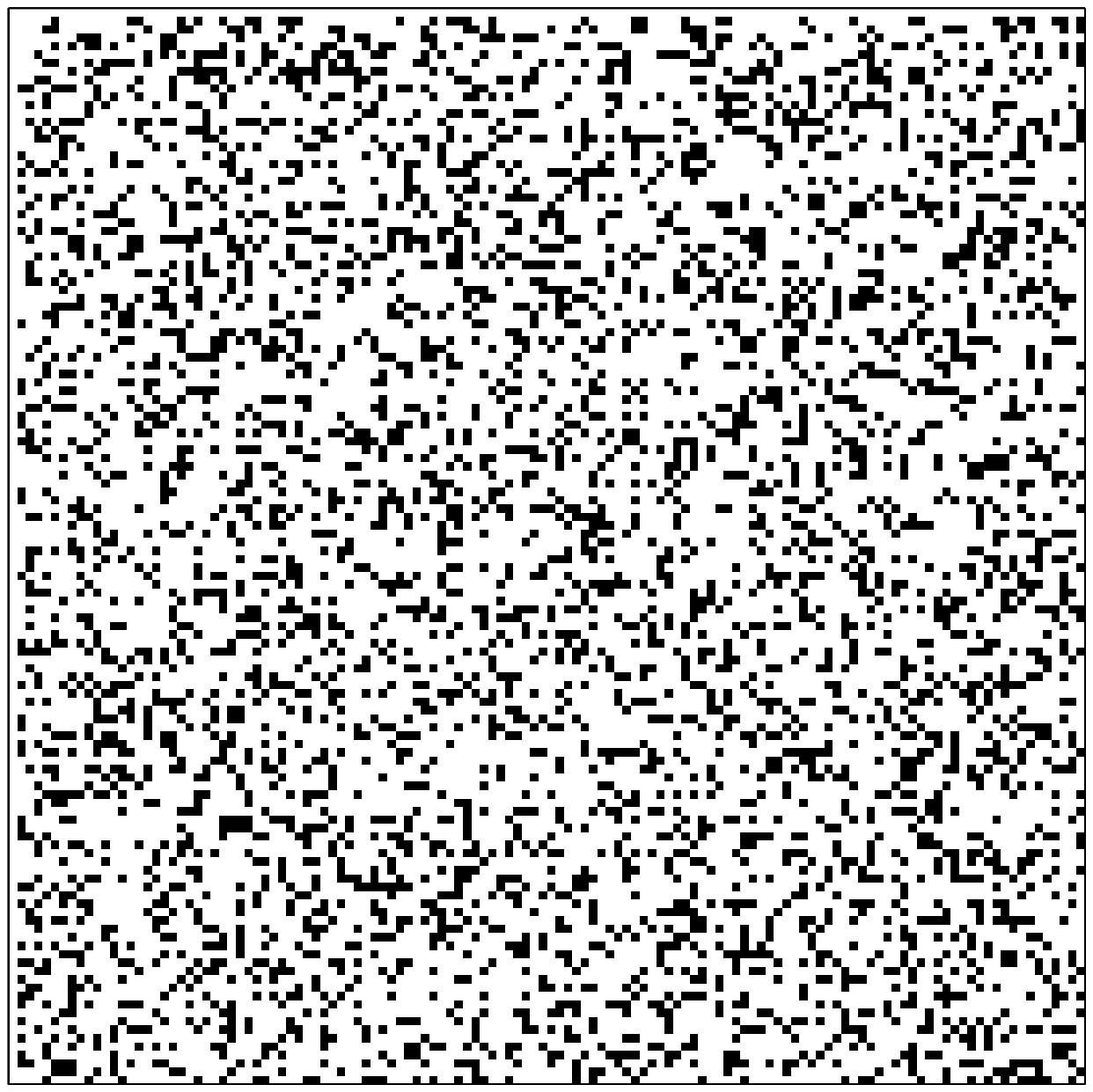}
\end{center}
\caption{\label{fig:obssetup} A map of the primary beam considered is represented in the left panel, with darkness increasing with the value of the function $A$ in the range [$0,1$]. A map of the mask made up of a random selection of  $25$ per cent of the spatial frequencies up to the band limit in the Fourier plane is represented in the right panel, with a unit value at the position of the selected frequencies identified by black points, and zero otherwise.}
\end{figure}
Aperture synthesis in radio interferometry is a powerful technique in radio astronomy dating back to more than sixty years ago \citep{ryle46,blythe57,ryle59,ryle60,thompson04}. In this context, the small portion of the celestial sphere around the pointing direction tracked by a radio telescope array during observation defines the original real planar signal or image probed $x$. The field of view observed is limited by a primary beam represented by a function $A$ with values in the range [$0,1$]. Each telescope pair at one instant of observation identifies a baseline defined as the relative position between the two telescopes. To each baseline is associated one measurement called visibility. In the simplest setting one also considers baselines with negligible component in the pointing direction of the instrument. Under this additional assumption, if the signal is made up of incoherent sources, each visibility corresponds to the value of the Fourier transform of the signal multiplied by the primary beam at a spatial frequency $\bm{k}$ identified by the components of the baseline in the plane of the signal.

We assume that the spatial frequencies $\bm{k}$ probed by all telescope pairs during the observation belong to the discrete grid of points $\bm{k}_{i}$ (see \citet{wiaux09a} for more details). The Fourier coverage provided by the $M/2$ spatial frequencies probed $\bm{k}_b$, with $1\leq b\leq M/2$, can simply be identified by a binary mask in the Fourier plane equal to $1$ for each spatial frequency probed and $0$ otherwise. The visibilities measured may be denoted by a vector of $M/2$ complex Fourier coefficients $\bm{y}\in\mathbb{C}^{M/2}\equiv\{y_{b}\equiv y(\bm{k}_b)\}_{1\leq b\leq M/2}$. Formally, the visibilities may equivalently be denoted by a vector of $M$ real measures $\bm{y}\in\mathbb{R}^{M}\equiv\{y_{r}\}_{1\leq r\leq M}$ consisting of the real and imaginary parts of the complex measures. In this discrete setting, the Fourier coverage is in general incomplete in the sense that the number of real constraints $M$ is smaller than the number of unknowns $N^2$: $M<N^2$. An ill-posed inverse problem is thus defined for the reconstruction of the signal, seen as a vector of real sampled values $\bm{x}\in\mathbb{R}^{N^2}\equiv\{x(\bm{p}_{i})\}_{1\leq i\leq N^2}$, from the measured visibilities $\bm{y}$. In the context of our analysis, the noise affecting the signal consists in the vector of real sampled values for the standard ${\rm CMB}$ components $\bm{g}\in\mathbb{R}^{N^2}\equiv\{g(\bm{p}_{i})\}_{1\leq i\leq N^2}$. The inverse problem therefore reads as:
\begin{eqnarray}
\bm{y}\equiv\mathsf{\Phi}\bm{x}+\bm{n} & \textnormal{ with } & \mathsf{\Phi}\equiv\mathsf{MFA} \nonumber\\
& \textnormal{ and with } & \bm{n} \equiv \mathsf{\Phi}\bm{g}, \label{eq:iip1}
\end{eqnarray}
where the measurement matrix $\mathsf{\Phi}\in\mathbb{C}^{(M/2)\times N^2}$ identifies the complete linear relation between the signal and the visibilities. The matrix $\mathsf{A}\in\mathbb{R}^{N^2\times N^2}\equiv\{A_{ij}\equiv A(\bm{p}_{i})\delta_{ij}\}_{1\leq i,j\leq N^2}$ is the diagonal matrix implementing the primary beam. The unitary matrix $\mathsf{F}\in\mathbb{C}^{N^2\times N^2}\equiv\{F_{ij}\equiv {\rm e}^{-{\rm i}\bm{k}_i\bm{\cdot}\bm{x}_j}/N\}_{1\leq i,j\leq N^2}$ implements the discrete Fourier transform providing the Fourier coefficients. The matrix $\mathsf{M}\in\mathbb{R}^{(M/2)\times N^2}\equiv\{M_{bj}\}_{1\leq b\leq M/2;1\leq j\leq N^2}$ is the rectangular binary matrix implementing the mask characterizing the interferometer. It contains only one non-zero value on each line, at the index of the Fourier coefficient corresponding to each of the spatial frequencies probed $\bm{k}_b$. Neglecting small correlations introduced between spatial frequencies by the primary beam, the power spectrum of the noise $\bm{g}$ is nonetheless modified as $P^{(Ag)}(k)=\vert \widehat{A} (k) \vert^2  P^{(g)}(k)$. The noise $\bm{n}\in\mathbb{C}^{M/2}\equiv\{n_{b}\}_{1\leq b\leq M/2}$ affecting the visibilities is Gaussian noise with different variances $P^{(Ag)}(k_b)$ for different components $n_b$. Let us recall that $\bm{g}$ is real so that the components of $\bm{n}$ are independent only when assuming that the mask $\mathsf{M}$ contains no pair of antipodal spatial frequencies $\{\bm{k_b},-\bm{k_b}\}$. For each such pair, the two corresponding noise components are complex conjugate of one another. In this work $M/2$ denotes the number of independent complex visibilities, while $M_0/2$, with $M_0\geq M$, will stand for the total number of complex visibilities including possible pairs of antipodal spatial frequencies.

\subsection{Whitening}

The measurement matrix may be augmented by a whitening operation which will be useful in a reconstruction perspective. It consists in dividing each measured visibility $y_b$ by the standard deviation of the real and imaginary parts of the corresponding noise component $n_b$. This operation is simply implemented in terms of the diagonal matrix $\mathsf{W}\in\mathbb{R}^{(M/2)\times (M/2)}\equiv\{W_{bb'}\equiv [P^{(Ag)}(k_b)/2]^{1/2}\delta_{bb'}\}_{1\leq b,b'\leq (M/2)}$. This gives rise to the modified inverse problem
\begin{eqnarray}
\tilde{\bm{y}}\equiv\tilde{\mathsf{\Phi}}\bm{x}+\tilde{\bm{n}} & \textnormal{ with } & \tilde{\mathsf{\Phi}}\equiv\mathsf{WMFA} \nonumber\\
& \textnormal{ and with } & \tilde{\bm{n}} \equiv \tilde{\mathsf{\Phi}}\bm{g}, \label{eq:iip1'}
\end{eqnarray}
where the whitened visibilities $\tilde{\bm{y}}\in\mathbb{C}^{M/2}\equiv\{\tilde{y}_{b}\equiv \tilde{y}(\bm{k}_b)\}_{1\leq b\leq M/2}$ are now affected by identically distributed independent (i.i.d.) Gaussian noise with unit variance of the real and imaginary part of each component $n_b$, i.e. white Gaussian noise.

For each test simulation, the visibilities are computed in accordance with relation (\ref{eq:iip1}). The primary beam $A$ considered for our observational set up is a Gaussian function with a full width at half maximum (${\rm FWHM}$) equal to the angular opening of the field of view ${\rm FWHM}=L$. The function has a maximum value equal to unity. Two Fourier coverages are considered independently, defined by masks made up of a random selection of $M_0$ spatial frequencies (see \citet{wiaux09a} for more details) with either $25$ per cent or $50$ per cent of the Fourier plane probed up to the band limit $B$. Note that, after discarding redundant visibilities associated with pairs of antipodal spatial frequencies in the two masks considered, these two percentages respectively correspond to selections of $M$ spatial frequencies with $21.7$ and $37.4$ per cent. The primary beam as well as the mask for a Fourier coverage of $25$ per cent are illustrated in Figure \ref{fig:obssetup}. The visibilities are also whitened in accordance with relation (\ref{eq:iip1'}) by the introduction of the suitable matrix $\mathsf{W}$ in each noise condition.
\subsection{Reconstruction basics} \label{sub:Reconstruction basics}
In the perspective of the reconstruction of the signal $\bm{x}$, relation (\ref{eq:iip1'}) represents the measurement constraint. Considering a candidate reconstruction $\bar{\bm{x}}$, the residual noise reads as $\bar{\bm{n}}\in\mathbb{C}^{M/2}\equiv\{\bar{n}_{b}\}_{1\leq b\leq M/2}\equiv\tilde{\bm{y}}-\tilde{\mathsf{\Phi}}\bar{\bm{x}}$. The residual noise level estimator, defined as twice the negative logarithm of the likelihood associated with $\bar{\bm{x}}$, simply reads as
\begin{equation}
\chi^{2}\left(\bm{y},\mathsf{\Phi},\bar{\bm{x}}\right)\equiv 2 \sum_{b=1}^{M/2}\frac{\big\vert  y_b - \left(\mathsf{\Phi}\bar{\bm{x}}\right)_b  \big\vert^2}{P^{(Ag)}\left(k_b\right)} \equiv \vert\vert \bar{n} \vert\vert_2^2 .\label{eq:iip2}
\end{equation}
The notation $\vert \cdot \vert$ for a scalar stands for the complex modulus when applied to a complex number and for the absolute value when applied to a real number. The  $\ell_{2}$ norm of the residual noise is the standard norm of the corresponding vector: $\vert\vert \bar{\bm{n}} \vert\vert_{2}\equiv(\sum_{b=1}^{M/2}\vert \bar{n}_{b}\vert^{2})^{1/2}$. This noise level estimator follows a chi-square distribution with $M$ degrees of freedom. Typically, this estimator should be minimized by the good candidate reconstruction. The measurement constraint on the reconstruction may be defined as a bound $\chi^{2}(\bm{y},\mathsf{\Phi},\bar{\bm{x}})\leq\epsilon^{2}$, with $\epsilon^{2}$ corresponding to some suitable large percentile of the $\chi^{2}$ distribution in order to avoid noise over-fitting.

In this context, many signals may formally satisfy the measurement constraint. A regularization scheme that encompasses enough prior information on the original signal is needed in order to find a unique solution. All image reconstruction algorithms will differ through the kind of regularization considered.

Firstly, the simplest possible reconstruction procedure would consist in assuming that the non-probed spatial frequencies have a null value, and simply inverting the Fourier transform after dividing each visibility by the variance of the noise in order to enhance the signal-to-noise ratio.  This reconstruction can by written as $\bm{x}_0 \propto (\mathsf{WMF})^{\dagger}\tilde{\bm{y}}$, where the notation $\mathsf{\cdot}^{\dagger}$ stands for the conjugate transpose operation. This map simply identifies with the convolution of the original signal and noise $\mathsf{A}(\bm{x}+\bm{g})$ by a beam $(\mathsf{F}^{\dagger}\mathsf{M}^{\dagger}\mathsf{W}^2\mathsf{M}\mathsf{F})\bm{\delta_0}$, where $\bm{\delta_0}$ stands for a spike at the center of the field of view: $\bm{\delta_0}\in\mathbb{R}^{N^2}\equiv\{\delta_{i_0j}\}_{1\leq j\leq N^2}$, with $i_0$ identifying the central pixel. In the standard vocabulary of radio interferometry, the beam is called the dirty beam, and the convolved map is called the dirty map in recognition of its visual appearance contrasting with the original signal. A normalization constant needs to be chosen so as to get a unit central value of the beam. In other words, no denoising or deconvolution is performed by this simple procedure that we call ${\rm INVERT}$. The only important operation in this reconstruction is the enhancement of the signal-to-noise ratio by the matrix $\mathsf{W}^2$.

Secondly, the standard ${ \rm CLEAN}$ algorithm is essentially a Matching Pursuit (${\rm MP}$) algorithm \citep{mallat93,mallat98} which performs the deconvolution by iterative removal of the dirty beam, thus leading from the dirty map to a clean map \citep{hogbom74,schwarz78,thompson04}. Formally, the algorithm should be stopped when the $\chi^{2}$ constraint above is satisfied.
\section{Compressed sensing reconstruction}\label{sec:Compressed sensing reconstruction}
In this section, we describe the standard global minimization problem ${\rm TV}_{\epsilon}$ for the reconstruction of the string signal in the context of compressed sensing. We model the prior statistical distribution of the gradient of the string signal with a ${\rm GGD}$, fitted on the basis of the training simulations. We also propose to enhance the ${\rm TV}_{\epsilon}$ problem on the basis of this statistical prior information, hence defining a ${\rm STV}_{\epsilon}$ minimization problem. We finally propose a power spectral model (PSM) used for independent estimation of the string tension, which may notably be useful for accelerating the convergence of the iterative algorithms solving the minimization problems.
\subsection{${\rm TV}_{\epsilon}$ problem}
For the reconstruction of a signal of sparsity $K$, the theory of compressed sensing proves that the number $M$ of random measurements required, in a sensing basis incoherent with the sparsity basis, roughly scales with the sparsity, well below the number of samples $N^2$ of the signal at Nyquist-Shannon rate: $M\propto K \ll N^2$ \citep{candes06b,candes06,candes08}. In the case of radio-interferometric data, the measurements are assumed to be random Fourier samples which represent a good sensing procedure for a string signal $x$ exhibiting sparse magnitude of the gradient $\nabla x$. In this case, in the presence of i.i.d. Gaussian noise, the so-called Total Variation minimization problem (${\rm TV}_{\epsilon}$) applies \citep{candes06a}, which corresponds to the minimization of the ${\rm TV}$ norm of the real signal under a $\chi^2$ constraint on the candidate reconstruction $\bar{\bm{x}}$:
\begin{equation}
\min_{\bar{\bm{x}}\in\mathbb{R}^{N^2}}\vert\vert\bar{\bm{x}}\vert\vert_{\rm TV}\textnormal{ subject to }\chi^{2}\left(\bm{y},\mathsf{\Phi},\bar{\bm{x}}\right)\leq\epsilon^2.\label{eq:csr1}
\end{equation}
The vector of sample values of the magnitude of the gradient associated with the vector $\bm{x}$ is denoted as $\bm{\nabla x}\in\mathbb{R}^{N^2}\equiv\{\nabla x(\bm{p}_{i})\}_{1\leq i\leq N^2}$. By definition, the ${\rm TV}$ norm of a signal is the $\ell_{1}$ norm of the magnitude of its gradient, i.e. $\vert\vert\bar{\bm{x}}\vert\vert_{\rm TV}=\vert\vert \bm{\nabla} \bar{\bm{x}}\vert\vert_{1}\equiv \sum_{i=1}^{N^2}\vert \nabla x(\bm{p}_{i})\vert$. We set $\epsilon^{2}$ to be the $99^{{\rm th}}$ percentile of the $\chi^{2}$ distribution.

The iterative algorithm considered for solving this convex optimization problem is based on the Douglas-Rachford splitting method \citep{combettes07, fadili09} in the framework of proximal operator theory \citep{moreau62}.
\subsection{Signal prior}
\begin{figure}
\begin{center}
\psfrag{xlabel}[c]{ \hspace{0.3cm}{\scriptsize $\nabla x$}}
\psfrag{ylabel}[c]{ \hspace{0.5cm}{\scriptsize $\log_{10}\mathcal{P}\left(\nabla x\vert\rho\right)$}}
\includegraphics[width=6cm,keepaspectratio]{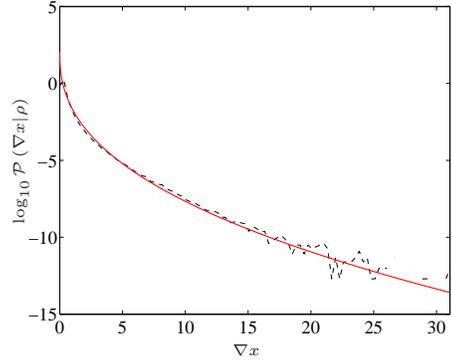}
\end{center}
\caption{\label{fig:histogram} Positive tail of modeled prior ${\rm GGD}$ distribution $\mathcal{P}(\nabla x\vert\rho)$ for a string tension $\rho=2\times 10^{-7}$. The ${\rm GGD}$ is superimposed on the histogram of the aggregate of the magnitude of the gradient of the $64$ test simulations.}
\end{figure}

In a Bayesian perspective, the minimization of the ${\rm TV}$ norm of the signal amounts to assuming that the signal gradient follows a Laplacian distribution with no spatial correlations, and to minimizing the negative logarithm of this prior distribution, i.e. to maximizing the distribution,  subject to the $\chi^{2}$ constraint. This promotes compressibility in a generic sense. A Laplacian distribution is indeed more peaked and more heavy-tailed than a Gaussian prior for example. However, with realistic simulations of the string signal at hand one can study more precisely the signal distribution. In that perspective, the signal gradient can be modeled at each sample point by the positive tail of a Generalized Gaussian Distribution (${\rm GGD}$):
\begin{equation}
\mathcal{P}\left(\nabla x\vert\rho\right)= \frac{q}{2\rho s \Gamma\left(q^{-1}\right)} {\rm e}^{-\vert\frac{\nabla x}{\rho s}\vert^{q}},\label{eq:csr2}
\end{equation}
where $\Gamma$ is the Gamma function, and where $q$ and $\rho s$ are respectively called shape and scale parameters. The shape parameter $q$ can be considered as a continuous measure of the compressibility of the underlying distribution. Setting $q=2$ recovers a Gaussian distribution, and $q=1$ recovers a Laplacian distribution. Letting $q$ approach $0$ yields compressible distributions. This parameter determines the kurtosis $\kappa^{(\nabla x)}$, i.e. the ratio of the fourth central moment to the square of the variance (second central moment), by
\begin{equation}
\kappa^{(\nabla x)}=\frac{\Gamma\left(5q^{-1}\right) \Gamma\left(q^{-1}\right)} {\left[\Gamma\left(3q^{-1}\right)\right]^{2}}.\label{eq:csr3}
\end{equation}
By nature this kurtosis is independent of the string tension $\rho$. The scale parameter $\rho s$ is linearly proportional to the standard deviation $\sigma^{(\nabla x)}(\rho)$ of the distribution. The corresponding variance reads as:
\begin{equation}
\left[\sigma^{(\nabla x)}\left(\rho\right)\right]^{2}= \rho^{2}\frac{\Gamma\left(3q^{-1}\right)}{\Gamma\left(q^{-1}\right)}s^{2}.\label{eq:csr4}
\end{equation}

The shape and scale parameters are estimated by a moment method on the aggregate of the magnitude of the gradient of the $64$ test simulations. The values with two significant figures are $\widehat{q}=0.42$ and $\widehat{s}= 2.3 \times 10^5$. The shape parameter value shows that the signal gradient exhibits a much more compressible gradient than a generic Laplacian distribution. Figure \ref{fig:histogram} shows the positive tail of the modeled prior ${\rm GGD}$ distribution $\mathcal{P}(\nabla x\vert\rho)$ for a string tension $\rho=2\times 10^{-7}$. The ${\rm GGD}$ is superimposed on the histogram of the aggregate of the magnitude of the gradient of the $64$ test simulations. Qualitatively, we see that the prior distribution is well modeled by a ${\rm GGD}$.
\subsection{${\rm STV}_{\epsilon}$ problem}
Accounting for the specific statistical prior information in the regularization of the inverse problem should enhance the reconstruction performance, beyond the standard ${\rm TV}_{\epsilon}$ minimization. In the Bayesian perspective discussed, we define a Statistical Total Variation ${\rm STV}_{\epsilon}$ minimization problem that consists in minimizing the logarithm of the ${\rm GGD}$ prior distribution modeled, subject to the $\chi^{2}$ constraint:
\begin{equation}
\min_{\bar{\bm{x}}\in\mathbb{R}^{N^2}}\vert\vert\bar{\bm{x}}\vert\vert_{\rm TV}^{\widehat{q}}\textnormal{ subject to }\chi^{2}\left(\bm{y},\mathsf{\Phi},\bar{\bm{x}}\right)\leq\epsilon^2.\label{eq:csr5}
\end{equation}
Again, we set $\epsilon^{2}$ to be the $99^{{\rm th}}$ percentile of the $\chi^{2}$ distribution. Because $\widehat{q}<1$ the problem is non-convex in nature, but a re-weighted scheme allows convergence of the algorithm through iterations of simple ${\rm TV}_{\epsilon}$ minimizations \citep{candes06a}.

Note that no prior knowledge of the string tension $\rho$ is formally required for the reconstruction. However we must acknowledge the fact that convergence appears to be faster when some prior estimation of $\rho$ is used to renormalize $\bar{\bm{x}}$ in the minimization term of relations (\ref{eq:csr1}) and (\ref{eq:csr5}), as expected from relation (\ref{eq:csr2}). As discussed in Section \ref{sec:Analyses and results}, such an estimation of $\rho$ will also define a threshold below which one should not expect good reconstruction of the string network. Our estimation procedure is described below.
\subsection{String tension estimation}\label{sec:String tension estimation}
By Bayes' theorem, the posterior probability distribution for $\rho$ given the measured visibilities $\mathcal{Q}(\rho\vert \bm{y})$ is simply obtained from the likelihood $\mathcal{L}(\bm{y}\vert\rho)$ and the prior probability distribution $\mathcal{P}(\rho)$ on $\rho$. For complete consistency, the likelihood should be calculated using the model established for the signal gradient and noise. However, while this model by construction accounts for the non-Gaussianity, i.e. sparsity of the string signal, it ignores the spatial correlations.

As in \citet{hammond09}, we have observed that a likelihood yielding a precise localization of the string tension value can actually be obtained using a power spectral model. Such a model assumes that both the string signal and the noise arise from statistically isotropic Gaussian random processes, such that their Fourier coefficients are independent Gaussian variables. Under this model, as the string signal and noise are independent, the observed signal $\mathsf{A}(\bm{x}+\bm{g})$ has a power spectrum, as modified by the primary beam, of the form:
\begin{equation}
P^{(A(x+g))}\left(k,\rho\right)=P^{(Ag)}\left(k\right)+P^{(Ax)}\left(k,\rho\right),\label{eq:csr6}
\end{equation}
with $P^{(Ag)}(k)$  defined in Section \ref{sub:Reconstruction basics} and similarly $P^{(Ax)}(k,\rho)=\vert \widehat{A} (k) \vert^2  P^{(x)}(k,\rho)$. In this setting the likelihood can be computed easily in terms of the measured visibilities $\bm{y}$ as: 
\begin{equation}
\mathcal{L}\left(\bm{y}\vert\rho\right) = \prod_{b=1}^{M/2} \frac{1}{\pi P^{(A(x+g))}\left(k_b,\rho\right)}{\rm e}^{-\frac{\vert y_b\vert^2}{P^{(A(x+g))}\left(k_b,\rho\right)}}.\label{eq:csr7}
\end{equation}
The posterior probability distribution for $\rho$ given the measured visibilities thus reads as 
\begin{equation}
\mathcal{Q}\left(\rho\vert\bm{y}\right)= D^{-1}\mathcal{P}\left(\rho\right)\mathcal{L}\left(\bm{y}\vert\rho\right),\label{eq:csr8}
\end{equation}
with normalization $D=\int \mathcal{P}(\rho)\mathcal{L}(\bm{y}\vert\rho) \: {\rm d}\rho$. We take the prior $\mathcal{P}(\rho)$ to be flat in an interval $\rho\in[0,\rho^{({\rm max})}]$, with an upper bound $\rho^{({\rm max})}$ large relative to the upper bound (\ref{eq:ssn4}).

We define an estimation $\widehat{\rho}$ of the string tension as the expectation value of this posterior probability distribution:
\begin{equation}
\widehat{\rho}=E\left[\mathcal{Q}\left(\rho\vert\bm{y}\right)\right].\label{eq:csr9}
\end{equation}
\section{Analyses and results}\label{sec:Analyses and results}
In this section, we firstly describe our numerical analyses, including  the performance criteria of reconstruction, and discuss their results. We also compare the corresponding eye visibility thresholds on the string tension with the detectability threshold obtained on the basis of the PSM. We finally propose a small discussion of our results.
\subsection{Analyses}
The overall reconstruction is effective for mapping the string network if the magnitude of gradient of the reconstructed signal closely resembles the magnitude of gradient of the true string signal. A simple qualitative measure of the performance is given by whether the string network is visible in the magnitude of gradient of the reconstructed signal. We define the eye visibility threshold as the minimum string tension around which the overall denoising and mapping by the magnitude of the gradient begins to exhibit string features visible by eye. We will augment this qualitative assessment of the reconstruction performance with three quantitative measures, namely the signal-to-noise ratio, the correlation coefficient and the kurtosis of the magnitude of the gradient of the reconstructed string signal re-multiplied by the primary beam. The kurtosis is known to be a good statistic for discriminating between models with and without cosmic strings \citep{moessner94}.

The signal-to-noise ratio is defined in terms of the original and reconstructed string signal gradients as:
\begin{equation}
{\rm SNR}^{(\nabla Ax,\nabla A\bar{x})}=-20\log_{10}\frac{\sigma^{(\nabla Ax-\nabla A\bar{x})}}{\sigma^{(\nabla Ax)}},\label{eq:r1}
\end{equation}
where
$\sigma^{(\nabla Ax-\nabla A\bar{x})}$ and $\sigma^{(\nabla Ax)}$ respectively stand for the standard deviations of the discrepancy signal $\nabla Ax-\nabla A \bar{x}$ and of the original signal $\nabla Ax$. The standard deviations are estimated from the sample variances on the basis of the signal realizations concerned. With this definition, the ${\rm SNR}$ is measured in decibels (${\rm dB}$). We will consider that the reconstruction is meaningful in terms of signal-to-noise ratio for the values of $\rho$ where this statistic is positive.

The correlation coefficient is also defined in terms of the original and reconstructed string signal gradients as:
\begin{equation}
r^{(\nabla Ax,\nabla A\bar{x})}= \frac{{\rm cov}^{(\nabla Ax,\nabla A\bar{x})}}{\sigma^{(\nabla Ax)}\sigma^{(\nabla A\bar{x})}},\label{eq:r2}
\end{equation}
where ${\rm cov}^{(\nabla Ax,\nabla A\bar{x})}$ stands for the covariance between $\nabla Ax$ and $\nabla A\bar{x}$, and $\sigma^{(\nabla Ax)}$ and $\sigma^{(\nabla A\bar{x})}$ respectively stand for the standard deviations of the signals $\nabla Ax$ and $\nabla A\bar{x}$. The covariance and variances are again estimated from the sample covariance and variances on the basis of the signal realizations concerned. We will consider that the reconstruction is meaningful in terms of correlation coefficient for the values of $\rho$ where this statistic is also positive. Note that the variances and covariances discussed obviously depend on the string tension, but not their dimensionless ratios. For this reason we have omitted this dependence in the definitions (\ref{eq:r1}) and (\ref{eq:r2}).

Analogously, kurtoses of the magnitude of the gradient of the reconstructed string signal re-multiplied by the primary beam are estimated from the sample kurtoses on the basis of the signal realizations concerned. The kurtosis of the magnitude of gradient of a string signal as expected on the basis of our test string signal simulation takes the value $\kappa^{(\nabla Ax)}=51$ with two significant numbers.

We compare the reconstruction results of the simple ${\rm INVERT}$ procedure and of the standard ${\rm CLEAN}$ algorithm with those of the  ${\rm TV}_{\epsilon}$ and ${\rm STV}_{\epsilon}$ algorithms. Reconstructions are performed for string tensions equi-spaced in logarithmic scaling in the range $\log_{10}\rho\in[-9,-5]$, corresponding to ratio values for $\rho$ of $1.0$, $1.6$, $2.5$, $4.0$, and $6.3$ in each order of magnitude. For each string tension, each algorithm, each noise condition (${\rm PA-IN}$, ${\rm SA-tSZ}$, or ${\rm SA+tSZ}$) and each coverage condition ($25$ or $50$ per cent) analysed, we consider that the quantitative measures described above indicate effective performance when this performance is significant over the $30$ corresponding simulations.
\subsection{Results}
\begin{table}
\begin{center}
\begin{tabular}{cccc}
\hline 
\noalign{\vskip\doublerulesep}
Noise & Coverage (per cent) & PSM detectability  & Eye visibility  \\
\hline
${\rm PA-IN}$ & $25$ & $7.8 \times 10^{-10}$ & $1.6 \times 10^{-9}$\\
${\rm PA-IN}$ & $50$ & $7.1 \times 10^{-10}$ & $1.0 \times 10^{-9}$\\ 
${\rm SA-tSZ}$ & $25$ & $2.9 \times 10^{-7}$ & $6.3 \times 10^{-7}$\\
${\rm SA-tSZ}$ & $50$ & $2.4 \times 10^{-7}$ & $4.0 \times 10^{-7}$\\
${\rm SA+tSZ}$ & $25$ & $5.0 \times 10^{-7}$ & $1.0 \times 10^{-6}$\\
${\rm SA+tSZ}$ & $50$ & $4.4 \times 10^{-7}$ & $6.3 \times 10^{-7}$\\
\hline
\end{tabular}
\end{center}
\caption{\label{tab:detectionthresholds} PSM detectability thresholds and eye visibility thresholds on the string tension, for each of the noise and coverage conditions considered. All values are given with two significant figures.}  
\end{table}
\begin{figure}
\begin{center}
\includegraphics[width=4cm,keepaspectratio]{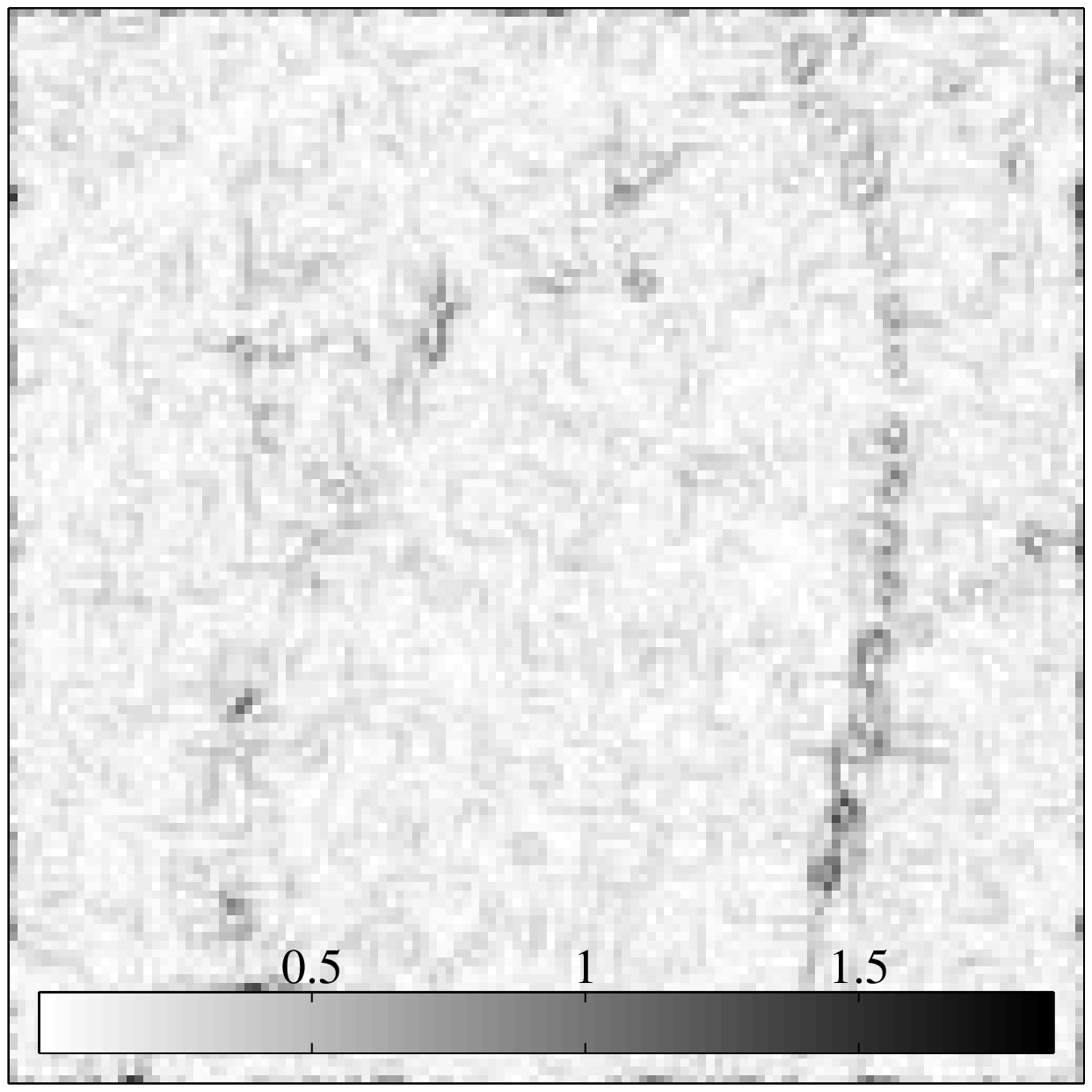}
\hspace{0.2cm}
\includegraphics[width=4cm,keepaspectratio]{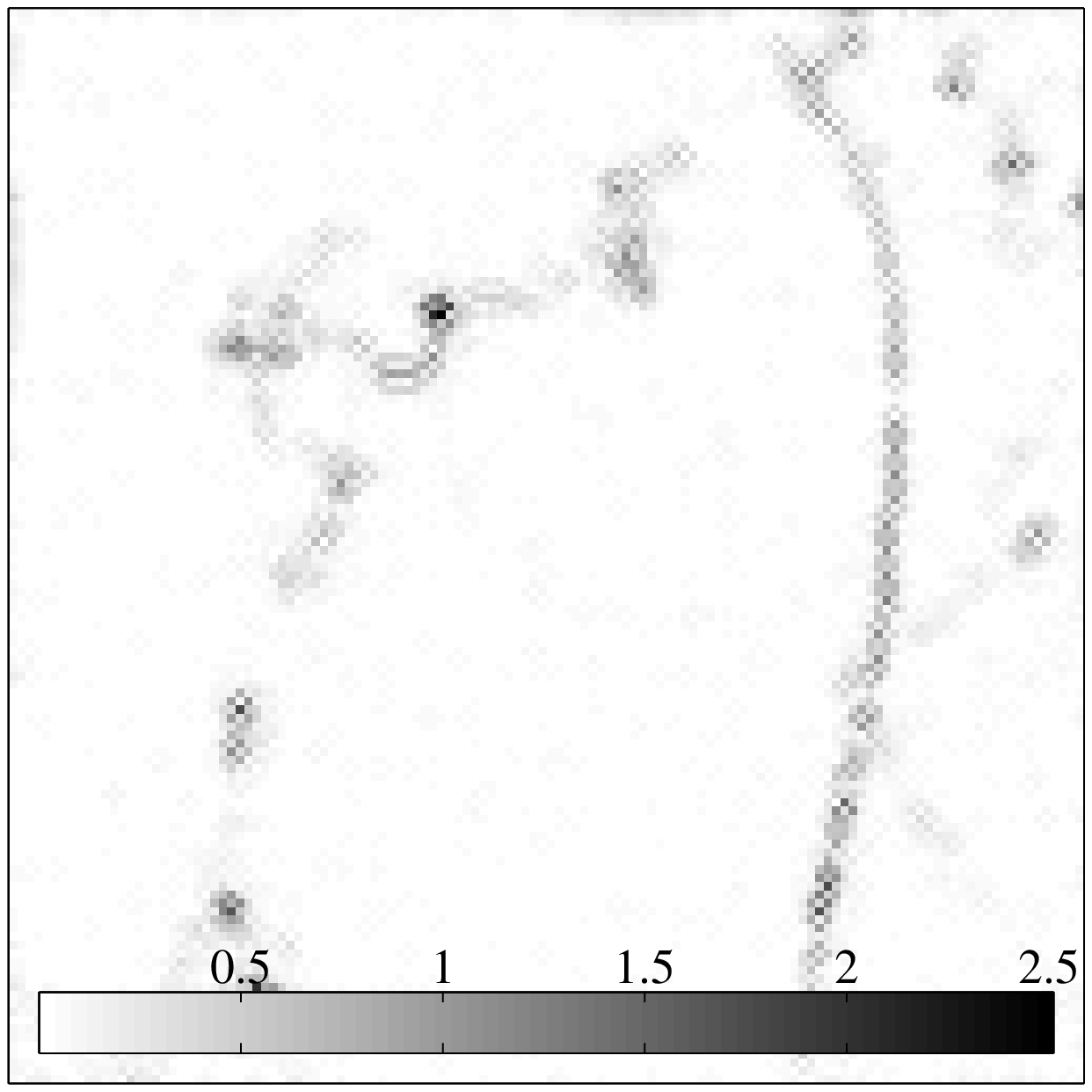}\\
\includegraphics[width=4cm,keepaspectratio]{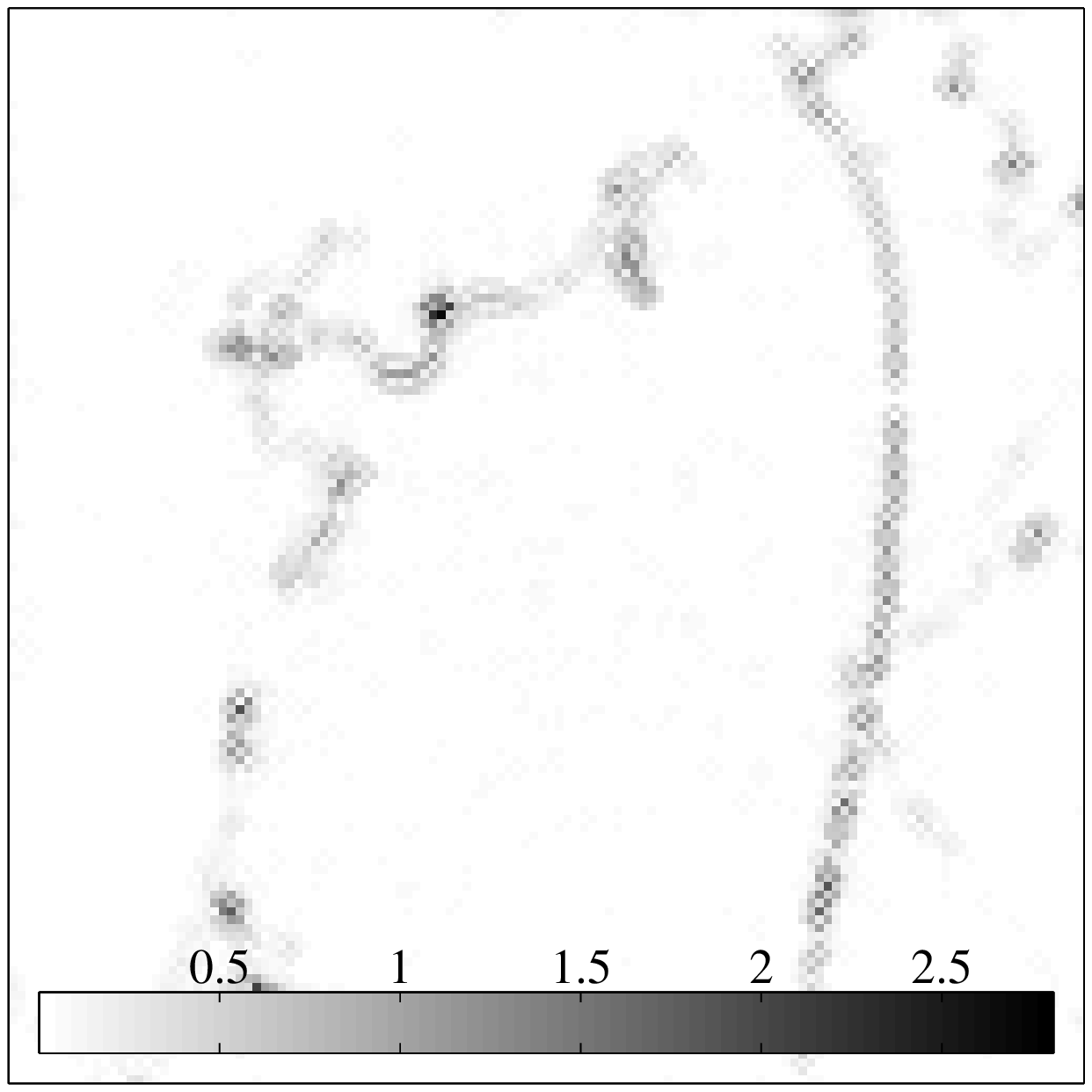}
\hspace{0.2cm}
\includegraphics[width=4cm,keepaspectratio]{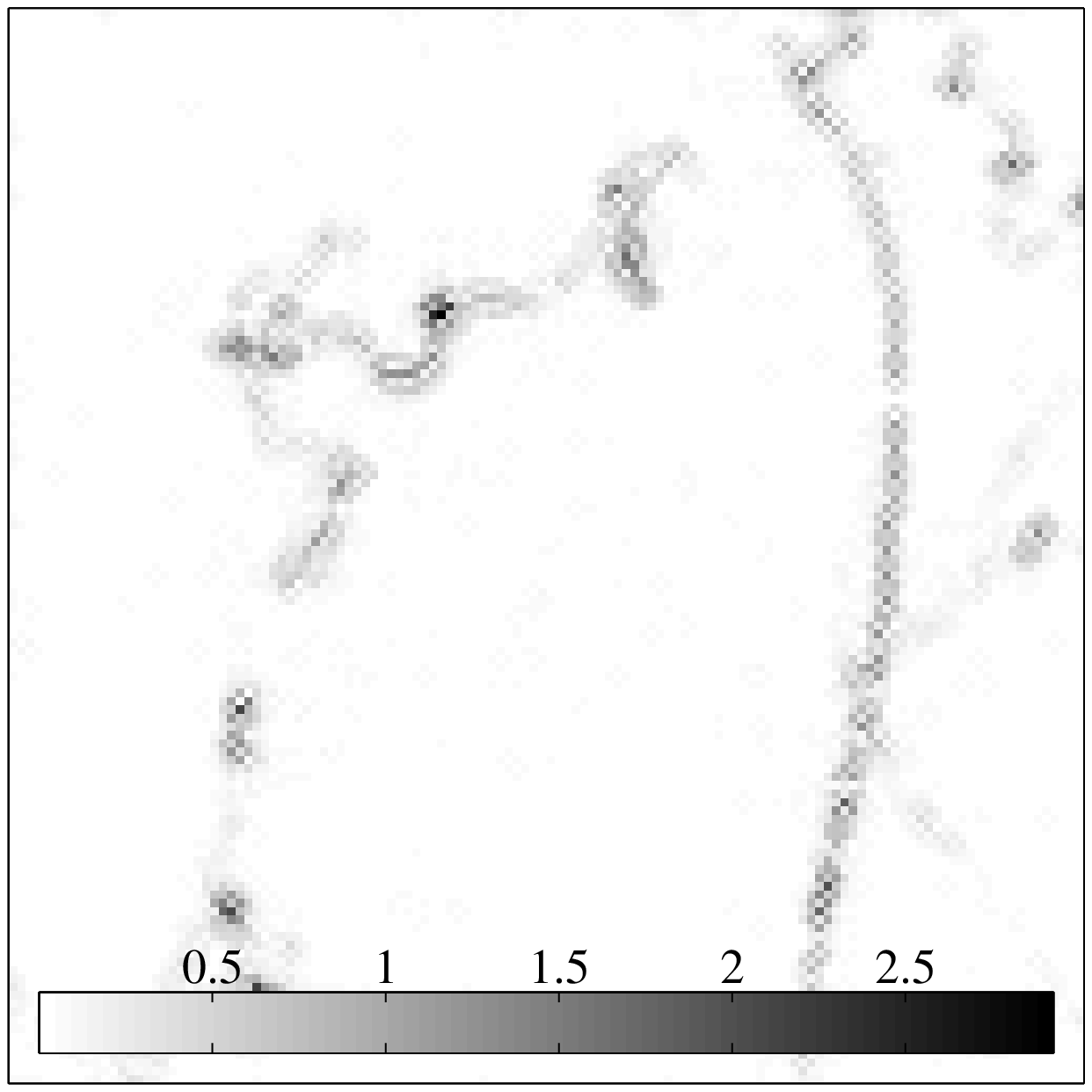}
\end{center}
\caption{\label{fig:PA-IN-rec} Magnitude of the gradient of reconstructed string signals re-multiplied by the primary beam, for a tension $\rho = 1 \times 10^{-7}$, in the noise condition ${\rm PA-IN}$ and for a coverage at $25$ per cent. The reconstructions considered are those associated with the simple ${\rm INVERT}$ procedure (top left panel), ${\rm CLEAN}$ (top right panel), ${\rm TV}_{\epsilon}$ (bottom left panel), and ${\rm STV}_{\epsilon}$ (bottom right panel).}
\end{figure}
\begin{figure*}
\begin{center}
\psfrag{CLEAN}[b]{ \hspace{0.6cm}{\tiny CLEAN}}
\psfrag{TV}[b]{ \hspace{0.6cm}{\tiny ${\rm TV}_{\epsilon}$}}
\psfrag{STV}[b]{ \hspace{0.6cm}{\tiny ${\rm STV}_{\epsilon}$}}
\psfrag{String tension}[]{ \hspace{0.8cm}{\scriptsize String tension}}
\psfrag{SNR (dB)}[b]{ \hspace{1cm}{\scriptsize ${\rm SNR}$ (${\rm dB}$)}}
\psfrag{Correlation}[b]{ \hspace{1.3cm}{\scriptsize Correlation coefficient}}
\psfrag{Kurtosis}[b]{ \hspace{1cm}{\scriptsize Kurtosis}}
\includegraphics[width=5cm,keepaspectratio]{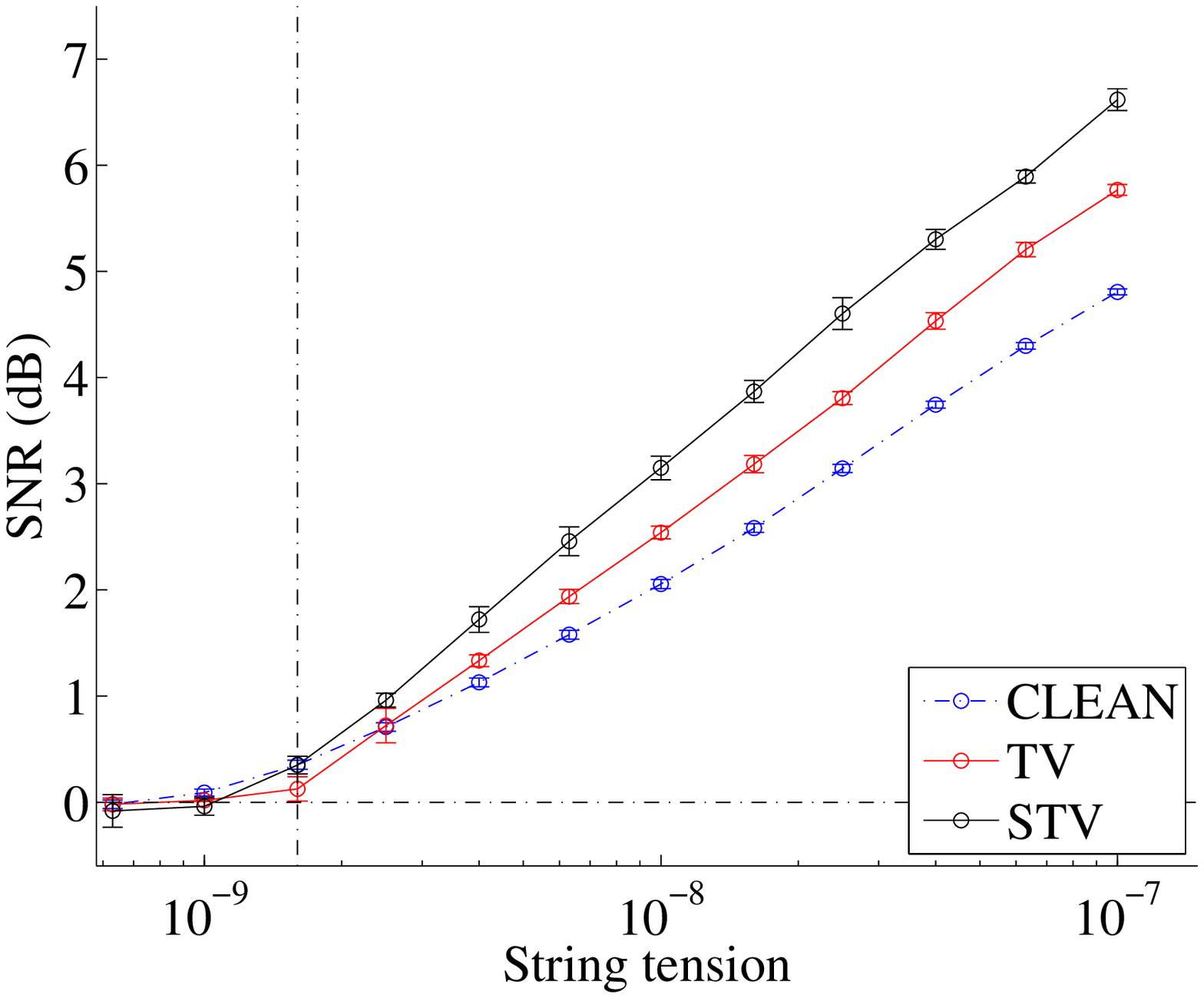}
\hspace{1cm}
\includegraphics[width=5cm,keepaspectratio]{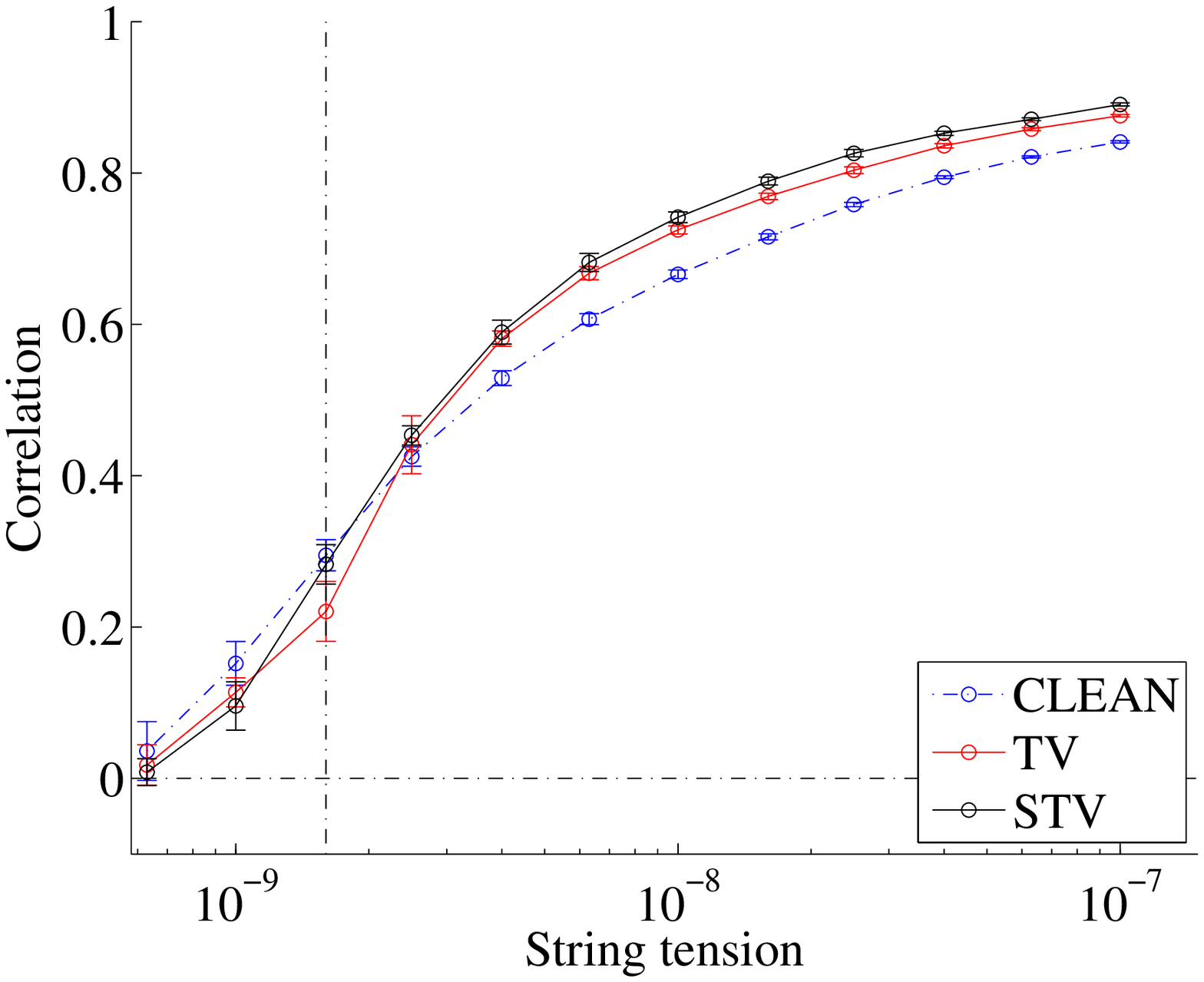}
\hspace{1cm}
\includegraphics[width=5cm,keepaspectratio]{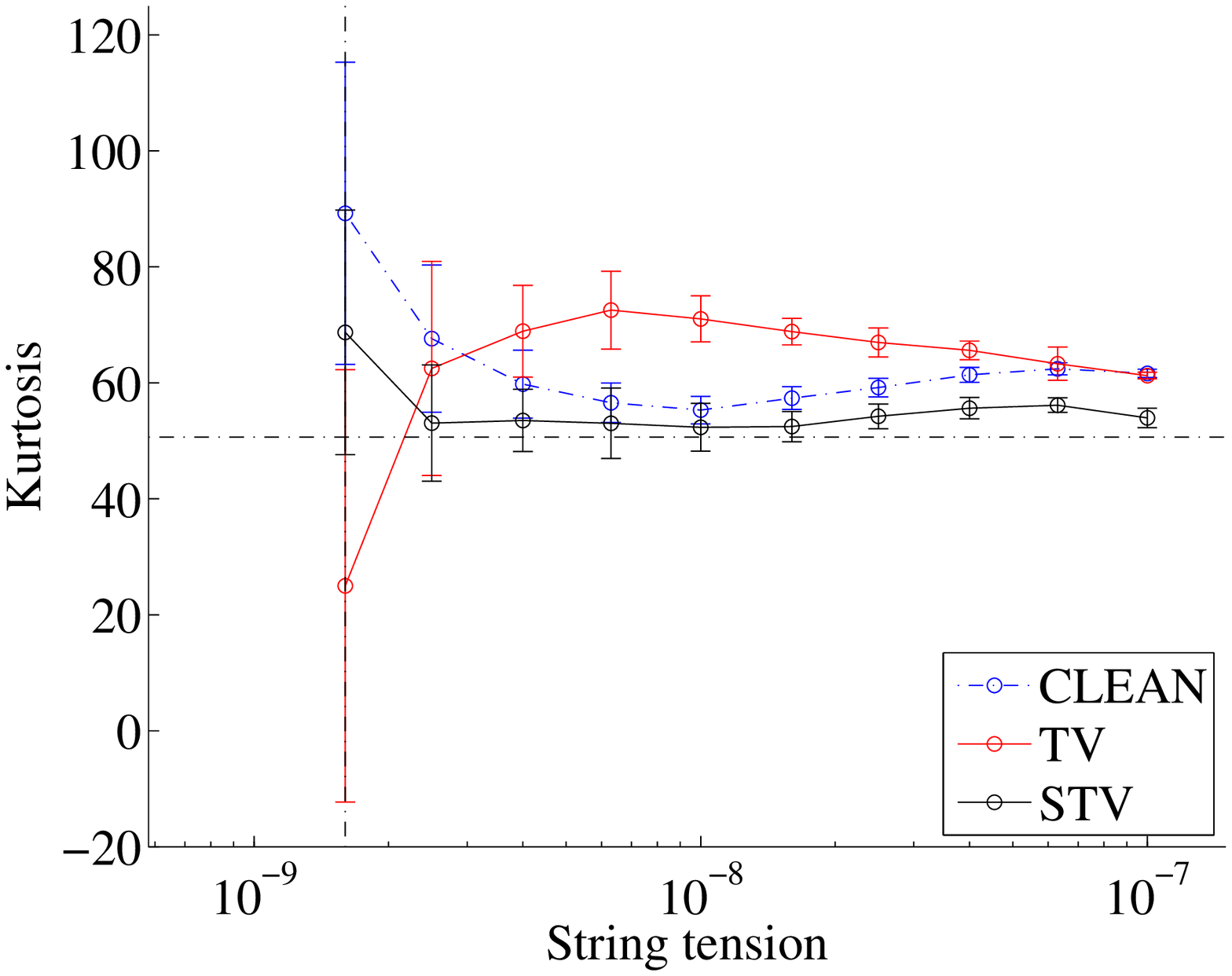}\\
\includegraphics[width=5cm,keepaspectratio]{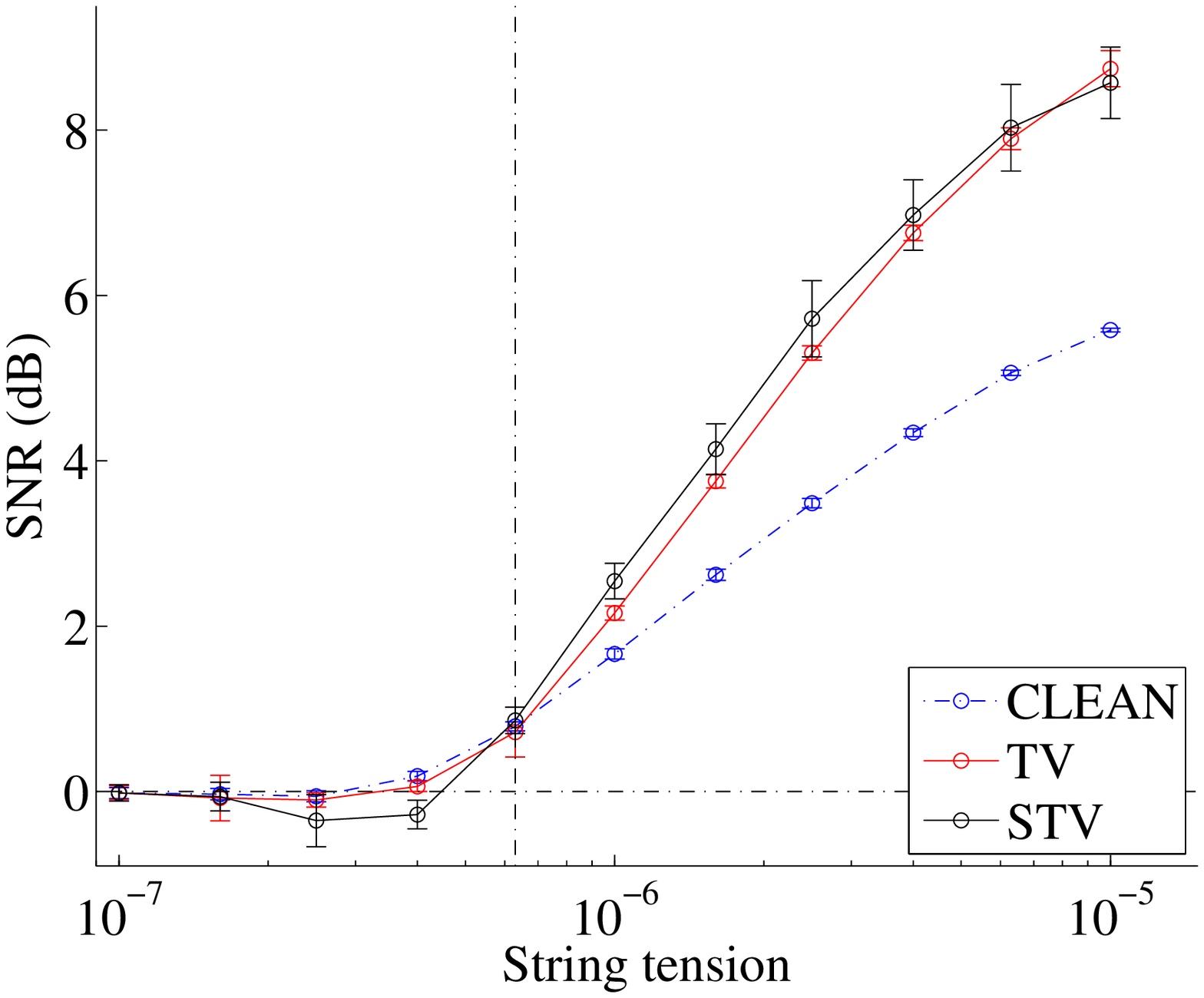}
\hspace{1cm}
\includegraphics[width=5cm,keepaspectratio]{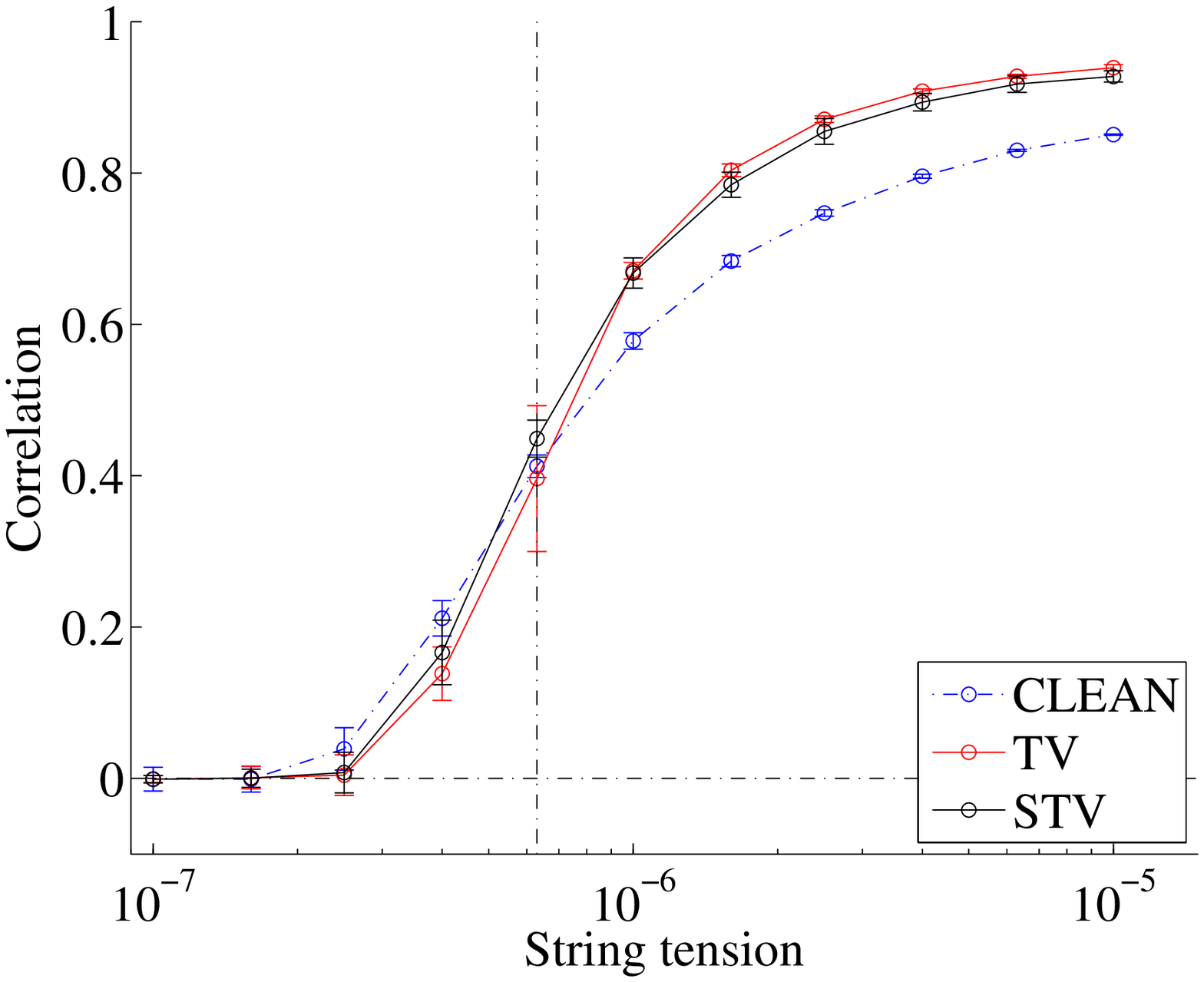}
\hspace{1cm}
\includegraphics[width=5cm,keepaspectratio]{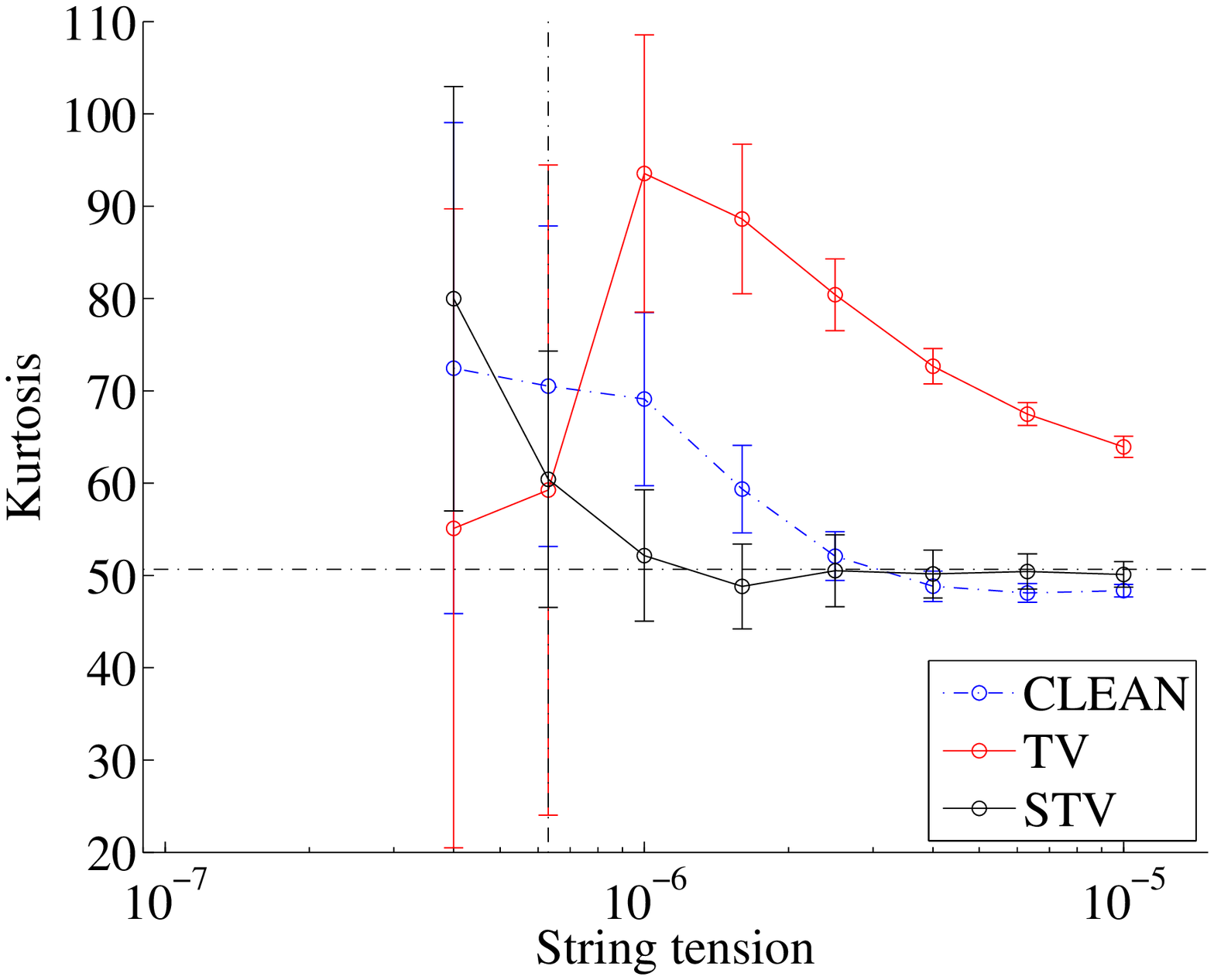}\\
\includegraphics[width=5cm,keepaspectratio]{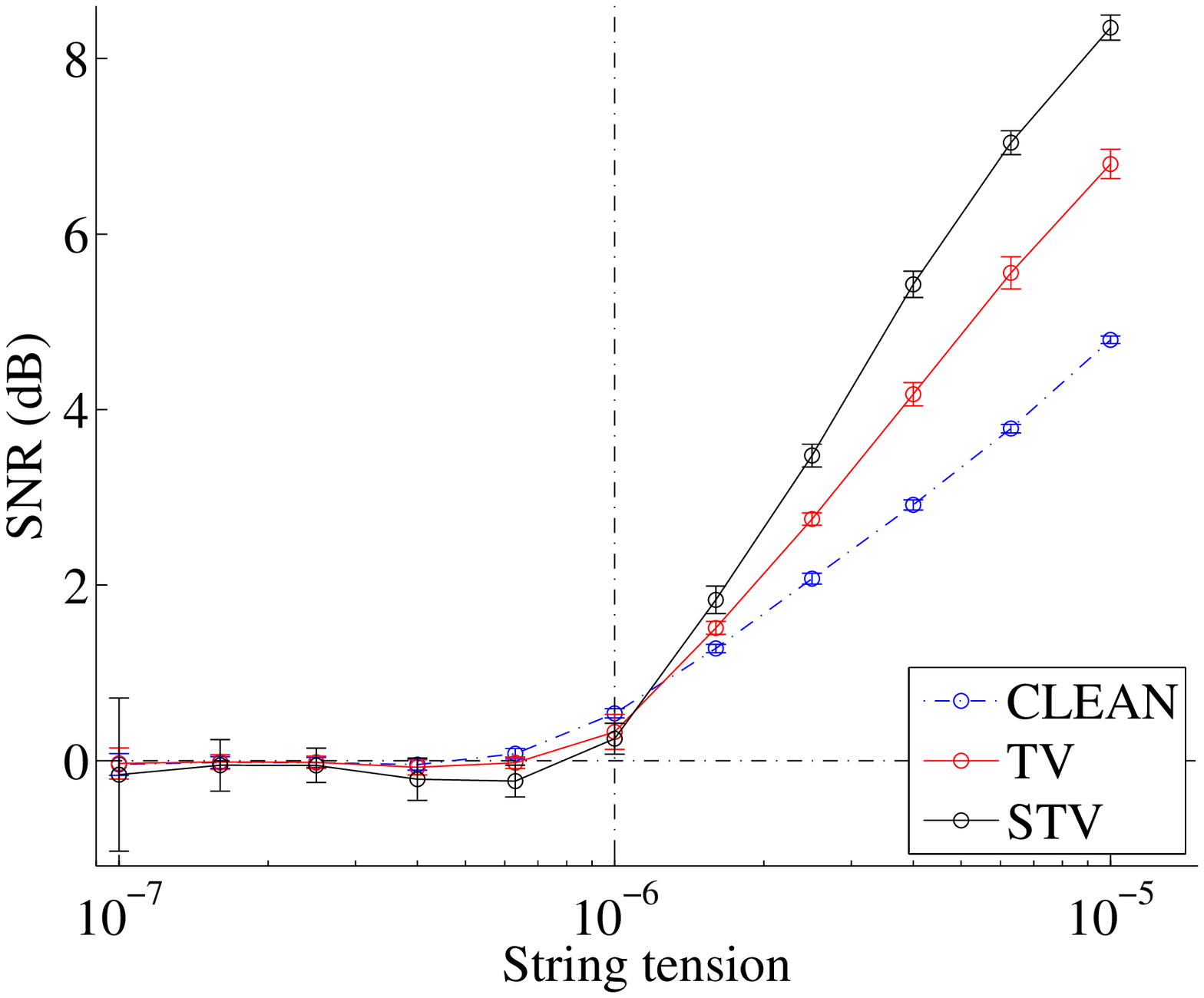}
\hspace{1cm}
\includegraphics[width=5cm,keepaspectratio]{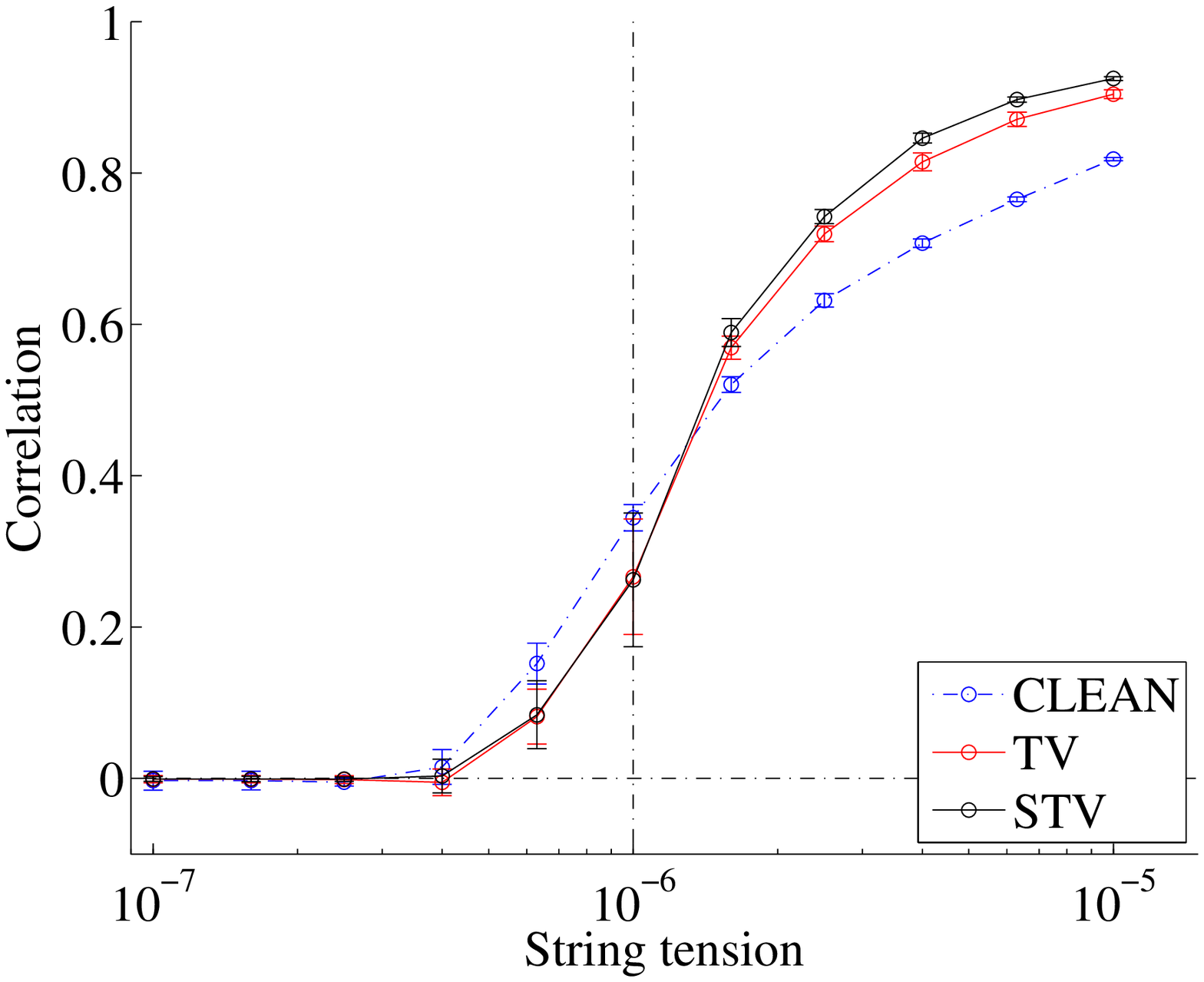}
\hspace{1cm}
\includegraphics[width=5cm,keepaspectratio]{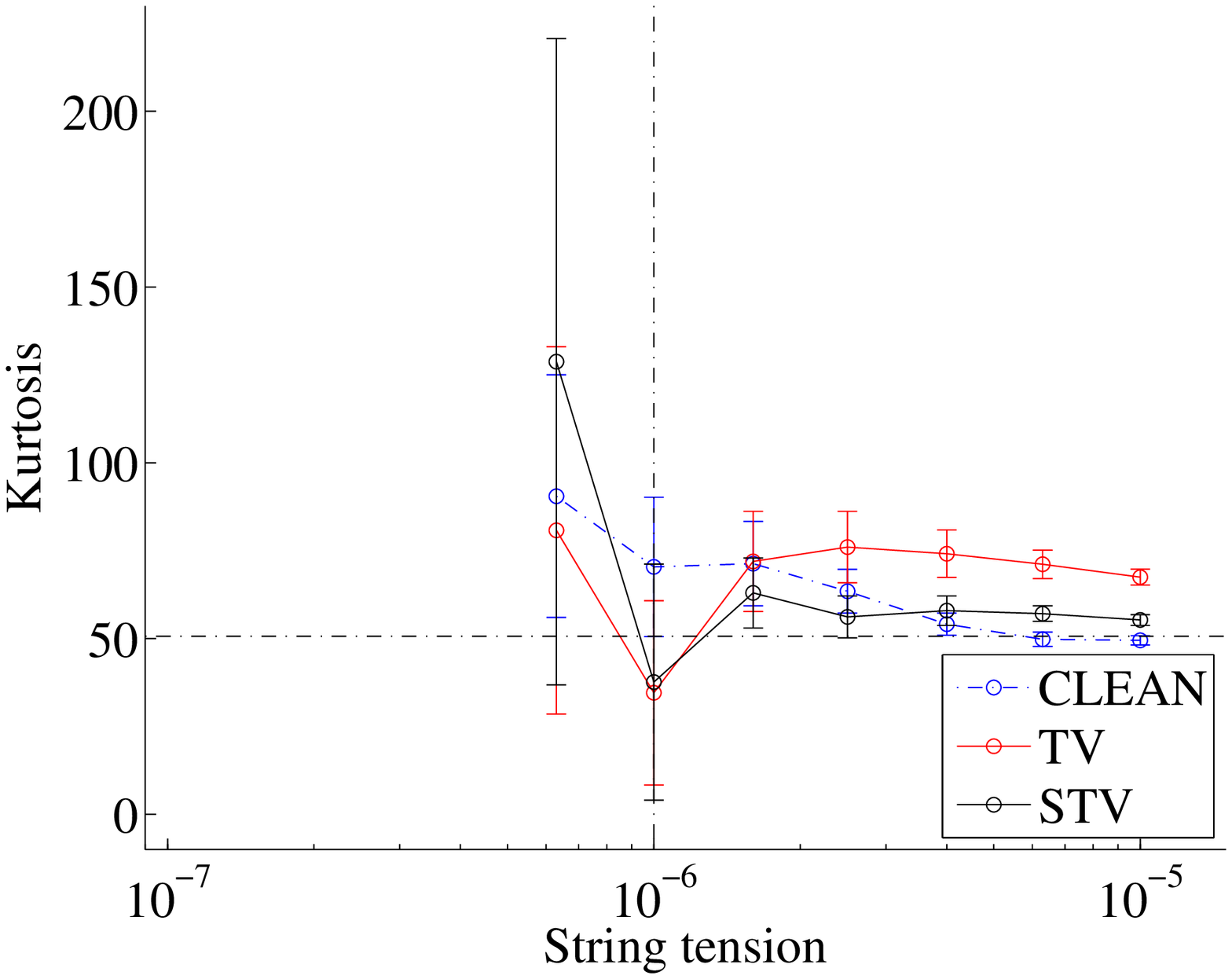}\\
\end{center}
\caption{\label{fig:stat} ${\rm SNR}$ in decibels (${\rm dB}$) (left panels), correlation coefficient (center panels), and kurtosis (right panels) of the magnitude of the gradient of reconstructed string signals re-multiplied by the primary beam as functions of the string tension in logarithmic scaling in the range $\log_{10}\rho\in[-9,-5]$, for a coverage at $25$ per cent in the noise conditions ${\rm PA-IN}$ (top panels), ${\rm SA-tSZ}$ (middle panels), and ${\rm SA+tSZ}$ (bottom panels). The reconstructions considered are those associated with ${\rm CLEAN}$, ${\rm TV}_{\epsilon}$, and ${\rm STV}_{\epsilon}$. The vertical lines on the curves represent the variability at one standard deviation of the estimated statistic across the $30$ test simulations considered (these lines are not visible where smaller than the width of the curves).  The black dot-dashed vertical lines represent the eye visibility thresholds. The black dot-dashed horizontal lines identify either the limit of zero ${\rm SNR}$, zero correlation coefficient, or the kurtosis of the magnitude of gradient of a string signal as expected from the test string simulation itself: $\kappa^{(\nabla Ax)}=51$.}
\end{figure*}

In all noise and coverage conditions, the simple ${\rm INVERT}$ procedure is already effective at enhancing the visibility of strings by eye thanks to the multiplication by $\mathsf{W}^2$ in Fourier space, but the resulting maps are dirty. The string tensions representing the eye visibility thresholds are listed in Table \ref{tab:detectionthresholds}. These thresholds are well below the experimental upper bound in the noise condition ${\rm PA-IN}$, while for the noise conditions ${\rm SA-tSZ}$ and ${\rm SA+tSZ}$ they are respectively of the same order and larger. This evolution is naturally due to the larger amount of noise at high spatial frequencies, where the signal gradient lives, in the presence of secondary anisotropies.

The eye visibility thresholds for ${\rm CLEAN}$, ${\rm TV}_{\epsilon}$, and ${\rm STV}_{\epsilon}$ are the same as for ${\rm INVERT}$. However it clearly appears that these more evolved algorithms offer much better reconstructions. This can already be acknowledged by simple eye inspection, as illustrated in Figure \ref{fig:PA-IN-rec} for the magnitude of the gradient of the reconstructed signals re-multiplied by the primary beam in the noise condition ${\rm PA-IN}$ at $25$ per cent and for a string tension $\rho=1\times 10^{-7}$. This qualitative assessment is confirmed by our quantitative measures for the reconstruction performance in all noise and coverage conditions and for all string tensions. Figure \ref{fig:stat}  represents the ${\rm SNR}$, correlation coefficient, and kurtosis of the magnitude of the gradient of reconstructed string signals re-multiplied by the primary beam as functions of the string tension in all noise conditions and for a coverage at $25$ per cent, for ${\rm CLEAN}$, ${\rm TV}_{\epsilon}$, and ${\rm STV}_{\epsilon}$. The corresponding curves for ${\rm INVERT}$ are actually not shown for readability reasons as they are very far from optimal values. In this context, the major interest of the analysis actually resides in the comparison of the standard ${\rm CLEAN}$ algorithm with the ${\rm TV}_{\epsilon}$ and ${\rm STV}_{\epsilon}$ algorithms.

Firstly, the reconstructions illustrated in Figure \ref{fig:PA-IN-rec} seem very similar for ${\rm CLEAN}$, ${\rm TV}_{\epsilon}$, and ${\rm STV}_{\epsilon}$. Note however that the magnitude of the gradient of the original signal (not shown on the figure) re-multiplied by the primary beam lies roughly in the range [$0,3.5$]. The range of amplitudes reported on the maps therefore indicate that the dynamic range of the reconstructions is better for ${\rm TV}_{\epsilon}$ than for ${\rm CLEAN}$, and even better for ${\rm STV}_{\epsilon}$. This is a generic behaviour that can be observed in all noise and coverage conditions and at all string tensions.

Secondly, as shown in Figure \ref{fig:stat} in all noise conditions and for a coverage at $25$ per cent, the eye visibility thresholds identify with the string tension where the ${\rm SNR}$ and correlation coefficient get down to very small values, and the kurtosis exhibits a large variability over the $30$ simulations around the expected value. This allows us to identify rigorously the eye visibility threshold as the string tension below which the algorithms are not effective anymore at recovering the string network. 

Thirdly, as also shown in Figure \ref{fig:stat} in all noise conditions and for a coverage at $25$ per cent, the ${\rm SNR}$ and correlation coefficient are always significantly larger for ${\rm TV}_{\epsilon}$ than for ${\rm CLEAN}$. They are always at least as good for ${\rm STV}_{\epsilon}$ as for ${\rm TV}_{\epsilon}$ and often significantly larger. However, it appears that the kurtosis is less optimally recovered with ${\rm TV}_{\epsilon}$ than with ${\rm CLEAN}$. But the ${\rm STV}_{\epsilon}$ approach regularizes the kurtosis significantly better than ${\rm CLEAN}$ in most cases.

Note that these last two conclusions also hold for a $50$ per cent coverage, hence suggesting their stability and robustness relative to the coverage. The reconstruction performance is always significantly better at $50$ per cent than at $25$ per cent, but the improvement is not drastic. This suggests that a small number of visibilities can already lead to good reconstructions of the string signal.
\subsection{PSM detectability threshold}
In the perspective of the evaluation of the performance of the reconstruction algorithms, it is natural to ask whether the requirement for a complete reconstruction of the signal strongly limits the lowest string tension accessible in comparison with the lowest detectable string tension on the basis of the PSM.

In this regard, for any possible string tension $\rho$, the probability distribution for the estimate $\widehat{\rho}$ on the basis of the PSM may be obtained from simulations. One can identify the critical value $\rho_{0}$ such that, for null string tension, the probability that the estimate $\widehat{\rho}$ is larger than  $\rho_{0}$ is small: $p(\widehat{\rho}\geq\rho_{0})=\alpha$, for some suitable positive value $\alpha$ much smaller than unity. The test for the hypothesis of null string tension is then defined as follows. For estimated values $\widehat{\rho}\geq\rho_{0}$, the hypothesis of null string tension may be rejected with a significance level $\alpha$. On the contrary for estimated values $\widehat{\rho}<\rho_{0}$, the hypothesis of null string tension may not be rejected.

We define the PSM detectability threshold $\rho^{\star}$ such that, for the string tension $\rho^{\star}$, the probability that the estimate $\widehat{\rho}$ is larger than  $\rho_{0}$ is large: $p(\widehat{\rho}\geq\rho_{0})=1-\beta$, for some suitable positive value $\beta$ much smaller than unity. Consequently, for string tensions larger than $\rho^{\star}$, the probability of rejecting a null string tension on the basis of the hypothesis test defined is larger than $1-\beta$, called the power of the test. The value $\rho^{\star}$ is the smallest string tension that can be discriminated from the hypothesis of null string tension for given values of significance and power of the test. It may thus be understood as a detectability threshold determined on the basis of the PSM.

The PSM detectability thresholds in the various noise and coverage conditions considered are reported in Table \ref{tab:detectionthresholds} for $\alpha\simeq\beta\simeq 0.01$. In the noise condition ${\rm PA-IN}$, detectability thresholds are well below the experimental upper bound, while for the noise conditions ${\rm SA-tSZ}$ and ${\rm SA+tSZ}$ they are respectively of the same order and larger. In all cases, the PSM detectability threshold $\rho^*$ is only slightly below the eye visibility threshold associated with the reconstruction algorithms discussed, by a factor smaller than $2$. This discrepancy between these two thresholds confirms that the detection problem alone can be solved with the PSM only down to slightly lower string tensions than the more difficult reconstruction problem. As already emphasized in \citet{hammond09}, it is one thing to estimate a single global parameter such as the string tension on the basis of a PSM, but quite another to explicitly reconstruct the string network itself. These comparison results are thus already very positive.
\subsection{Discussion}
Our results strongly support the idea of the superiority of the compressed sensing approaches for a string signal reconstruction from radio-interferometric data, in particular when specific statistical prior information is introduced in the global minimization problem defined for reconstruction. However, at the high angular resolution considered and in the noise condition ${\rm SA+tSZ}$, the eye visibility thresholds, as well as the PSM detectability thresholds, are above the current best experimental upper bound on the string tension. One first approach to circumvent this difficulty in the perspective of the reconstruction of a possible string network from forthcoming radio-interferometric observations is to consider an observation frequency around $217$ ${\rm GHz}$ so as to eliminate the thermal ${\rm SZ}$ effect and to fall in the noise condition ${\rm SA-tSZ}$. But even in this case the eye visibility thresholds remain slightly higher than the experimental upper bound.

In this context, recent work by the authors might help to drastically enhance the signal reconstruction performance in the context of compressed sensing approaches \citep{wiaux09b}. Radio interferometers with small field of view and baselines with non-negligible and constant component in the pointing direction should be considered. In this context, the visibilities measured essentially identify with a noisy and incomplete Fourier coverage of the product of the planar signal multiplied by the primary beam with a linear chirp modulation. Theoretical and numerical results show the universality of the corresponding spread spectrum phenomenon relative to the sparsity basis, in terms of the achievable quality of reconstruction through global minimization problems such as ${\rm TV}_{\epsilon}$ and ${\rm STV}_{\epsilon}$. Preliminary numerical results regarding the application of such a framework to the reconstruction of a string signal in the ${\rm CMB}$ look very promising.

In another line of thoughts, global component separation techniques might be envisaged in order to simultaneously extract all non-Gaussian components of the ${\rm CMB}$ temperature data, including the ${\rm SZ}$ effects and the string signal.

Finally, let us recall that the string signal simulations used for modelling the prior statistical distribution of the signal gradient as ${\rm GGD}$'s and for computing its power spectrum are realistic. Nevertheless, further analysis should be performed in order to study the stability of the ${\rm STV}_{\epsilon}$ reconstruction quality relative to a discrepancy between the statistical distribution of the true signal and our model. Analyzing the difference between the ${\rm TV}_{\epsilon}$ and ${\rm STV}_{\epsilon}$ reconstruction qualities already provides useful information in that respect, as the ${\rm TV}_{\epsilon}$ minimization problem captures the essence of the compressiblity of the signal gradient, but with a non-exact ${\rm GGD}$ shape parameter $q=1$. In this context, one can argue in favour of a relative stability of the reconstruction quality in terms of the ${\rm SNR}$ and correlation coefficient as these measures are indeed smaller for ${\rm TV}_{\epsilon}$ than for ${\rm STV}_{\epsilon}$, but still larger than for ${\rm CLEAN}$. This stability is lost in terms of the kurtosis though.
\section{Conclusion} \label{sec:Conclusion}
The ${\rm STV}_{\epsilon}$ algorithm proposed for the reconstruction of a string signal from arcminute resolution radio-interferometric data of the ${\rm CMB}$ is designed in the framework of compressed sensing. It notably relies on a model of the prior statistical distribution of the signal, fitted on the basis of realistic simulations. The algorithm shows superior performance relative to the standard ${\rm CLEAN}$ algorithm. Secondary ${\rm CMB}$ anisotropies strongly hamper the signal reconstruction though, and further work is still required in order to obtain good performance at string tensions below the current experimental upper bound.
\section*{Acknowledgments}
The authors wish to thank M. J. Fadili for private communication of results on optimization by proximal methods. The authors also thank A. A. Fraisse, C. Ringeval, D. N. Spergel, and F. R. Bouchet for kindly providing simulations of a string signal. The authors also thank the reviewer for his valuable comments.
 Y. W. is Postdoctoral Researcher of the Belgian National Science Foundation (F.R.S.-FNRS).

\label{lastpage}

\end{document}